\documentclass[12pt,preprint]{aastex}
\usepackage{url}
\usepackage[usenames]{color}
\usepackage{amsmath}
\usepackage{lineno}

\newcommand\gta{\lower 0.5ex\hbox{$\buildrel > \over \sim\ $}} 
\newcommand\lta{\lower 0.5ex\hbox{$\buildrel < \over \sim\ $}} 

\begin{document}

\title{Stark-Broadened Profiles for Ionized Helium Lines Using Computer Simulations}

\author{Patrick Tremblay, Alain Beauchamp, and Pierre Bergeron}
\affil{D\'epartement de Physique, Universit\'e de Montr\'eal,
  C.P.~6128, Succ.~Centre-Ville, Montr\'eal, Qu\'ebec H3C 3J7, Canada}
\email{patrick@astro.umontreal.ca, bergeron@astro.umontreal.ca}

\begin{abstract}
We present new and improved calculations of Stark-broadened
  profiles for ionized helium, a key ingredient in the spectroscopic
  analysis of helium-atmosphere DO white dwarfs. Our approach builds
  upon the computer simulation framework previously developed for
  neutral helium, which fully accounts for the dynamical interactions
  of both ions and electrons with the emitting helium atom.  We extend
  this theoretical formalism by relaxing the assumption of
  straight‑line trajectories for the perturbing particles (electrons
  and ionized helium) and adopting the hyperbolic trajectories
  appropriate for their interaction with a charged emitter, thereby
  accounting for their dynamical influence on the line‑broadening
  process.  In this exploratory study, we focus on the He {\sc ii}
  $\lambda$4686 line, the strongest absorption feature observed in the
  spectra of DO white dwarfs. We present the resulting Stark profiles
  and perform a detailed comparison with those available in the
  literature.
\end{abstract}


\section{Introduction} \label{intro}

White dwarf stars represent the endpoint of more than 97\% of the
stars in the Galaxy, and the determination of their properties thus
provides a variety of useful information on the star formation history
in our galaxy, and on stellar evolution in general
\citep{Saumon2022}. White dwarfs are found in a variety of flavors:
The DA stars, whose spectra are dominated by hydrogen lines, comprise
about 80\% of the white dwarf population, and their atmospheres are
presumably hydrogen-rich. The remaining 20\% are generally referred to
as non-DA stars. These include the DB (DO) stars, whose spectra are
dominated by neutral (ionized) helium lines, the DQ stars that show
molecular carbon Swan bands, the DZ stars with metallic absorption
features, and the DC stars, which are completely featureless. A
significant fraction of white dwarf stars enter the cooling sequence
with helium-dominated atmospheres (DO-type) but eventually develop
hydrogen-dominated atmospheres (DA-type) due to the upward diffusion
of residual hydrogen in their envelope -- the so-called float-up model
(see \citealt{Bedard2024} for a review).

\citet{Bedard2020} empirically studied this phenomenon by conducting a
detailed spectroscopic analysis of 1806 hot white dwarfs ($T_{\rm eff} \geq
30,000$ K) selected from the Sloan Digital Sky Survey (SDSS, \citealt{York2000}). Their
study demonstrated that approximately 24\% of white dwarfs begin their
evolution as DO stars, of which about two-thirds subsequently
transform into DA stars. The atmospheric parameters derived from their
model atmosphere analysis are summarized in the $\log g - T_{\rm eff}$ diagram
shown in Figure 13 of \citet{Bedard2020}, where several peculiar
behaviors can be observed. On the one hand, at low effective
temperatures, most DA and DO/DB white dwarfs form a continuous
sequence, essentially parallel to constant-mass curves and tightly
centered around $M \sim 0.55$ $M_\odot$, in agreement with our
understanding of white dwarf evolution. On the other hand, at high
temperatures, DA stars with $T_{\rm eff} \gtrsim 60,000$ K exhibit
lower-than-average surface gravities, while DO stars with $T_{\rm eff} \gtrsim
50,000$ K display higher-than-average values. Consequently, the
hydrogen-dominated sequence bends upward toward lower masses, whereas
the helium-dominated sequence sharply drops toward higher
masses. These features are in clear contradiction with the
well-established facts that white dwarfs evolve at a constant mass and
that the DA and DB spectral classes at lower temperatures have indeed
similar mean masses \citep{Tremblay2019,Genest2019a}. These mass
discrepancies are not unique to the study of
\citet{Bedard2020}. Indeed, based on their compilation of
spectroscopic parameters of DO stars from the SDSS, \citet[][see their
  Figure 5]{Reindl2014} also obtained a mass distribution peaking at a
larger value of $M \sim 0.675$ $M_\odot$. They argue that this high-mass
peak is physically real, but the striking contrast between the mass
distributions of DO and DB stars suggests otherwise. Based on these
observations, \citet{Bedard2020} thus concluded that the spectroscopic
masses of very hot SDSS white dwarfs are affected by widespread issues
that are not limited to atmospheric models, cooling sequences, or
analysis techniques. One possible issue is the line broadening theory
of ionized helium (He {\sc ii}) used in model spectra for white
dwarfs.

All model spectra for hot white dwarfs published in the literature
rely on the Stark broadening calculations of \citet[][see also
  \citealt{Schoening89b}]{Schoening89} for the He {\sc ii} lines, in
particular for He {\sc ii} $\lambda$4686, which is the strongest
helium line observed in DO white dwarfs, and DAO stars as well. These
calculations are based on the Unified Theory of \citet{Vidal70}, but
including the effects of the ionic charge on the paths of the
electrons. Recently, we have used computer simulations to study the
Stark broadening of He {\sc i} lines \citep{Tremblay2026}, in order to
replace the semi-analytical calculations of \citet{Beauchamp97}
commonly used in spectroscopic analyses of DB white dwarfs and hot subdwarfs as well (see, e.g.,
\citealt{Voss2007,Bergeron2011,Koester2015,Latour2026,Heber2026}). It is in this context that we have undertaken
a similar investigation of the line broadening of ionized helium using
numerical simulations to verify the validity of the profiles of
\citet{Schoening89} employed in previous spectroscopic studies.  In
this particular work, we focus only on He {\sc ii} $\lambda$4686.  In
contrast with our previous computer simulations \citep{Tremblay2026}, the perturbed
emitting particle here is He {\sc ii}, which, unlike neutral helium,
is a charged particle. In this case, the perturbing particles
(electrons and He {\sc ii}) no longer follow straight-line
trajectories, a complex problem that we address in this paper.

We first introduce in Section 2 the fundamental concepts underlying
the calculation of Stark-broadened line profiles using computer
simulations. In Section 3, we review the method employed for these
calculations when the perturbing particles follow straight-line
trajectories and extend this approach to hyperbolic trajectories in
Section 4. The results of our He {\sc ii} $\lambda$4686 calculations
are presented in Section 5, followed by our conclusions in Section 6.

\section{Computer Simulations of Stark-Broadened Line Profiles}

\subsection{Extending Trivial Molecular Dynamics to Charged Emitters and Hyperbolic Trajectories}

The aim of this work is to generalize the computer simulation method
developed by several authors \citep{Stamm84, Gigosos87, Hegerfeldt88,
  Gigosos96, Gigosos2009,Lara2012,GomezPhD,Cho2022,Tremblay2026} to
calculate Stark-broadened profiles for the He {\sc ii} $\lambda$4686
line. Computer simulation is the state-of-the-art approach for
detailed Stark-broadened profile calculations, and tables of line
profiles of hydrogen and helium atoms have been calculated using a
version of the method called Trivial Molecular Dynamics (which is
described below).  In this method, the charged particles interacting
with the emitting (neutral) ion follow straight-line trajectories at
constant velocity, and adapting it to perturbers with more general
trajectories in the presence of a charged emitter requires targeted
modifications.

\citet{Poquerusse96} were the first to examine the applicability of the simulation method to charged emitters, adopting a hyperbolic orbit for the perturbers. 
They derived the general equation of the trajectory and proposed a numerical method to determine the position of a particle as a function of time along its trajectory. This approach involves inverting the non-linear equation that defines the hyperbolic eccentric anomaly as a function of time.
However, their work does not provide a statistical distribution of phase space coordinates allowing one to generate particles within the simulation volume at initialization, nor to handle incoming particles during the course of the simulation.

Approximately two decades ago, numerical simulation methods diverged into two main families. The first generation of simulation methods, the so‑called Trivial Molecular Dynamics (TMD) models treat the perturbers as if they experience no electrostatic interaction other than with the emitter located at the center of the simulation volume, and the electric field at the position of the emitter is computed with a screened-debye potentiel in order to approximate correlation effects in the particle's motion. 

In the case of an uncharged emitter, the trajectory reduces trivially to a straight line within TMD. However, when the central particle carries a charge, two levels of approximation can be employed, depending on the form of the adopted interaction potential. A pure Coulomb potential results in a hyperbolic orbit, whereas the more rigorous alternative of using a screened Debye potential---the same potential employed to compute the electric field in the vicinity of the emitter---leads to a more complex trajectory that cannot be expressed in analytical form. \citet{Stambulchik2006} followed this second, i.e. more rigourous, approach to generate line shapes of some H- and He-like ions.
While the TMD approach has become increasingly realistic for hydrogenic emitters—benefiting from major advances in the underlying quantum calculations, such as the use of a full Coulomb interaction rather than a multipole-expansion approximation and improved treatment for penetrating collisions \citep{Stambulchik2022}—no simulation-based line-profile tables for ionized helium are yet available for astrophysical applications.  

The second family of simulations, Full Molecular Dynamics (FMD), reproduces the physical processes more faithfully by adopting a Coulomb interaction between all particles, whether perturbers or the emitter, and by solving the equations of motion simultaneously for the entire system \citep{Gigosos2018}. This constitutes an $N$-body problem  that requires significantly more computing ressources than the TMD approach.

With the aim of generating line-profile tables for ionized helium, we investigate a TMD approach whose level of physical realism lies between the semi-analytical Stark-broadening calculations \citep{Schoening89} and computer  simulations such as those implemented by \citet{Stambulchik2006}. 
In accordance with the TMD methodology, the motion of the perturbers is determined solely by their interaction with the emitter, which in this work is described by a Coulomb potential rather than a Debye-screened one. This choice is less rigorous physically, but it allows the displacement of a particle along the resulting hyperbolic orbit to be modeled using a strategy similar to that proposed by \citet{Poquerusse96}.

\subsection{Quantum Treatment of the Emitting Ion }

The simulation method developed here preserves the essential elements of the TMD framework, including the correct initialization of perturbers within the simulation volume and the handling of new particles entering the domain over time. It differs, however, by replacing the traditionally employed autocorrelation function with the power spectrum, as proposed by \citet{Rosato2020} and \citet{Cho2022}.

The power spectrum is defined as  
\begin{equation}
P(\omega) \propto \omega^4
\lim_{T\to +\infty}  {1\over T} \sum_{ab} \rho_a
\Big< \Big| \int_{-T/2}^{T/2} dt \langle b | {\bf d}(t)|a \rangle e^{i\omega t} \Big|^2  \ \big>_{\rm av}
\end{equation}  
where $|a\rangle$ and $|b\rangle$ are the upper and lower states of the emitting ion for the considered transition, $\rho_a$ is the statistical weight of the upper states, and ${\bf d}(t)$ is the dipole operator of the emitting ion in the Heisenberg picture. The above equation includes a thermal average over all possible configurations of moving perturbers.  

As usual, the slowly varying $\omega^4$ term is excluded from the detailed calculations, yielding the line profile  
\begin{equation}
L(\omega) \propto 
\lim_{T\to +\infty}  {1\over T} \sum_{ab} \rho_a
\Big< \Big| \int_{-T/2}^{T/2} dt \langle b | {\bf d}(t)|a \rangle e^{i\omega t} \Big|^2\ \Big>_{\rm av}
\label{eqn:Lomega}
\end{equation}  
which is then normalized to unit area. In the simulation approach, the thermal average is approximated by an average over a large but finite set of $N$ dynamic configurations. The line profile is thus computed as the mean of $N$ individual profiles $L_i(\omega)$, each corresponding to a specific dynamic configuration $i$,  
\begin{equation}
L(\omega) = {1\over N} \sum_{i = 1}^N L_i(\omega)
\label{eqn:Lomega2}
\end{equation}  
where  
\begin{equation}
L_i(\omega) \propto 
\lim_{T\to +\infty}  {1\over T} \sum_{ab} \rho_a \Big| \int_{-T/2}^{T/2} dt \langle b | {\bf d}(t)|a \rangle e^{i\omega t} \Big|^2
\label{eqn:Liomega}
\end{equation}  
is computed from the $i$th temporal sequence of dipole moments.  

The no-quenching approximation is used in this work. Consequently, the matrix element of the dipole operator is expressed as an expansion over two disjoint sets of upper and lower states, $|a'\rangle$ and $|b'\rangle$, which include all states sharing the same principal quantum number (and spin) as the $|a\rangle$ and $|b\rangle$ states involved in the transition:  
\begin{equation}
    \langle b|{\bf d}(t)|a\rangle = \sum_{a'b'} \langle b|U_b^\dag(t)|b'\rangle 
    \langle b'|{\bf d}|a'\rangle \langle a'|U_a(t)|a\rangle
    \label{eqn:bda}
\end{equation}  
where ${\bf d}$ is now the dipole operator in the Schrödinger picture, and $U_{a,b}$ are the evolution operators projected onto the upper and lower subspaces of states, respectively.  

The projected evolution operators satisfy  
\begin{eqnarray}
i\hbar {dU_a(t)\over dt} &=& H_a(t)\  U_a(t) \nonumber \\
i\hbar {dU_b(t)\over dt} &=& H_b(t) \ U_b(t)
\label{eqn:UaUb}
\end{eqnarray}  
where $H_{a,b}$ are projections of the Hamiltonian  
\begin{equation}
H = H_0 + V(t)
\label{eqn:H}
\end{equation}  
onto the two relevant subspaces. Here, $H_0$ is the Hamiltonian of the unperturbed ion, while $V(t)$ represents the time-dependent interaction between the emitting ion and the moving perturbers. In this work, we consider only the first nonvanishing term in the multipole expansion of the interaction—namely, the dipole term:  
\begin{equation}
V(t) = e{\bf F}(t) \cdot {\bf R}
\end{equation}  
where ${\bf R}$ is the position operator of the radiating electron of the emitting ion.  

In each temporal sequence, the ion emitter remains at the center of a sphere, referred to as the simulation volume, with a radius on the order of the Debye radius. The analysis of the variations in the electric field ${\bf F}(t)$ experienced by the emitter focuses on contributions from the charged particles—ions and electrons—within the simulation volume. The moving perturbers are treated as independent classical particles, following trajectories determined by their charge and that of the central emitter. 

The line profile $L_i(\omega)$ associated with the $i$th simulation volume is computed in three steps. First, a time sequence of electric fields, ${\bf F}_i(t_k)$, at the origin of the simulation volume is generated based on the spatial distribution of moving perturbers for a series of $N_t$ discrete times, $t_k \equiv k\Delta_t$. Next, the Schrödinger equation (\ref{eqn:UaUb}) is numerically integrated. The integration scheme ensures that the evolution operator matrices remain unitary, and the time step $\Delta_t$ is chosen to be sufficiently small to resolve variations in the field \citep{Gigosos96,Gigosos2009,Gomez2016,Cho2022,Tremblay2026}. Finally, the time sequence of dipole moment matrix elements (equation \ref{eqn:bda}) is substituted into equation (\ref{eqn:Liomega}), where the integral over $t$ is performed for different frequencies.  

A key challenge of the computer simulation method is ensuring that, across all simulation volumes, the statistical properties of the system remain unchanged both over time and across  parallel simulations. For straight-line trajectories at constant velocity, this is achieved by choosing an appropriate coordinate system in six-dimensional phase space and deriving the corresponding joint statistical distribution. This distribution is used for both the initialization and reinjection phases of the simulation. Particle positions are then generated within this coordinate system, called the impact coordinate system \citep{Barnard69}, using a uniform random number generator. However, this system is not suitable for Stark broadening calculations involving charged emitters.  

To address this limitation, the proposed method introduces a coordinate system, referred to as the hyperbolic impact coordinate system. This framework ensures that generated particles follow hyperbolic trajectories while preserving the statistical properties of the system over time. Section \ref{sec:line} examines the case of an uncharged central particle with perturbers following straight-line trajectories, following the standard approach described in the literature on TMD. This involves replacing Cartesian or radial coordinates (${\bf r,v}$) in phase space with a six-dimensional coordinate vector that accounts for the trajectory shape. As will be shown, this second coordinate system (the hyperbolic impact system) is particularly well-suited for initializing particles within a sphere and randomly generating additional particles that enter and cross the sphere during the simulation. The complete set of equations governing particle generation with hyperbolic trajectories is detailed in Section \ref{sec:hyperb}.  

\section{Straight-Line Trajectory: Standard Method of the TMD \label{sec:line}}

This section presents the particle generating process relevant for the well-known case of a neutral emitter and charged perturbers following straight-line trajectories at constant velocity in the TMD model. This topic has been frequently addressed in the literature \citep{Stamm84,Gigosos87,Hegerfeldt88,Gigosos96,Gomez2016,Cho2022,Tremblay2026}, but a complete review of the steps involved in constructing the straight line trajectory and deriving the relevant joint statistical distributions in the impact coordinate system is necessary to familiarize oneself with the notation and more complex calculations required for hyperbolic trajectories.

\subsection{Parameterization of a Straight-Line Trajectory}

For a neutral emitter positioned at the center of the simulation volume, each perturbing particle follows a straight-line trajectory given by  

\begin{equation}
{\bf r}(t) = {\bf b} + {\bf v}(t - \tau_b)
\label{eqn:rlin}
\end{equation}

\noindent
where ${\bf v}$ is the constant velocity vector of the particle, $\tau_b$ is the time of closest approach, and ${\bf b}$ is the trajectory’s closest point to the origin. The vector ${\bf b}$ is perpendicular to ${\bf v}$, and its norm, $b$, represents the impact parameter. The velocity vector ${\bf v}$ is defined as  

\begin{equation}
{\bf v} \equiv v {\bf v_3}(\theta, \phi),
\end{equation}

\noindent
where $v$ is the magnitude of ${\bf v}$ and ${\bf v_3}$ is a unit vector. The angles ($\theta, \phi$) define the orthonormal basis:  
\begin{eqnarray}
{\bf v_1}&=&(\cos\phi \cos\theta, \sin\phi \cos\theta, -\sin\theta)^\text{T}\nonumber \\
{\bf v_2}&=&(-\sin\phi , \cos\phi, 0)^\text{T} \nonumber \\
{\bf v_3}&=&(\cos\phi \sin\theta, \sin\phi \sin\theta, \cos\theta)^\text{T},
\label{eqn:v1230} 
\end{eqnarray}

\noindent
where $0 \leq \phi < 2\pi$, $0 \leq \theta \leq \pi$, and $\text{T}$ denotes the transpose operation. When $\theta = \phi = 0$, these vectors align with the positive $x$, $y$, and $z$ axes, respectively. The perpendicularity condition between ${\bf b}$ and ${\bf v}$ implies that ${\bf b}$ can be expressed as a linear combination of ${\bf v_1}$ and ${\bf v_2}$:
\begin{equation}
{\bf b} = b(\cos \alpha {\bf v_1} + \sin \alpha {\bf v_2}),
\end{equation}

\noindent
introducing a new angle $\alpha$ in the range $0 \leq \alpha < 2\pi$.  

The six parameters ${\bf x} \equiv (\theta, \phi, \alpha, v, b, \tau_b)$ define a valid coordinate system, as each point (${\bf r, v}$) in phase space corresponds uniquely to a set of values ${\bf x}$ for any constant $t$. The only exception occurs at the origin ($|{\bf r}| = 0$), where the angular coordinates become degenerate. However, such points are not considered valid positions for perturbing particles in the simulation.  

The probability density function $p({\bf r, v})$, associated with an infinitesimal phase-space volume ${\bf d^3r\ d^3v}$, must align with statistical mechanics:  

\begin{equation}
p({\bf r}, {\bf v})   \propto  \ e^{-E({\bf r,\ v})/kT},
\label{eqn:pxv}
\end{equation}

\noindent
where $E$ is the total energy of the particle (kinetic plus potential), $k$ is Boltzmann’s constant, and $T$ is the temperature.  

To express this probability density in terms of the new coordinate system, we use the jacobian $J$ of the transformation:

\begin{equation}
p({\bf x}) \propto  \ e^{-E({\bf x})/kT}  \ J({\bf x}),
\label{eqn:pxv3}
\end{equation}

\noindent
where $E$ and  
\begin{equation}
J \equiv \bigg|{{\bf \partial ^3r \ \partial ^3v} \over{\bf \partial^6x}}\bigg|
\end{equation}

\noindent
are functions of ${\bf x}$. The jacobian is given by the determinant of the jacobian matrix:  
\begin{equation}
{\bf J} \equiv \left (
\begin{matrix}
{\partial {\bf r^\text{T}} \over \partial v} & {\partial {\bf v^\text{T}} \over \partial v} \\
{\partial {\bf r^\text{T}} \over \partial \theta} &{\partial {\bf v^\text{T}} \over \partial \theta}  \\
{\partial {\bf r^\text{T}} \over \partial \phi} &{\partial {\bf v^\text{T}} \over \partial \phi}  \\
{\partial {\bf r^\text{T}} \over \partial \tau_b} &{\partial {\bf v^\text{T}} \over \partial \tau_b}  \\
{\partial {\bf r^\text{T}} \over \partial b} &{\partial {\bf v^\text{T}} \over \partial b}  \\
{\partial {\bf r^\text{T}} \over \partial \alpha} &{\partial {\bf v^\text{T}} \over \partial \alpha}  
\end{matrix}
\right ),
\label{eqn:Jstraighline}
\end{equation}

\noindent
where the elements of the $6\times 6$ matrix are written as transposed (i.e., horizontal) 3D-vectors to reduce clutter in the equation.
\noindent
Because each partial derivative can be expanded as a linear combination of ${\bf v_{1,2,3}}$, the jacobian matrix factorizes into the product of two simpler $6\times6$ matrices:

\begin{equation}
{\bf J} = \left (
\begin{matrix}
0 &0 & (t-t_0) & 0 & 0 & 1 \\
v(t-t_0) & 0 & -b\cos\alpha & v & 0 & 0  \\
-b\sin\alpha & b\cos\alpha\cos\theta + v\sin\theta (t-t_0) &0 &0 &v\sin\theta  &0 \\
0 &0 & -v  &  0 &0& 0  \\
\cos\alpha & \sin\alpha & 0 & 0 & 0 & 0  \\
-b\sin\alpha & b\cos\alpha & 0 &  0 & 0& 0 
\end{matrix}
\right )
\left (
\begin{matrix}
{\bf v_1^\text{T}} & {\bf 0^\text{T}} \\
{\bf v_2^\text{T}} & {\bf 0^\text{T}} \\
{\bf v_3^\text{T}} & {\bf 0^\text{T}} \\
{\bf 0^\text{T}} & {\bf v_1^\text{T}} \\
{\bf 0^\text{T}} & {\bf v_2^\text{T}} \\
{\bf 0^\text{T}} & {\bf v_3^\text{T}} 
\end{matrix}
\right ),
\end{equation}
where ${\bf 0^\text{T}}$ are 3D-vectors with null cooordinates. Because the determinant of the second matrix is unity, the jacobian simplifies to the determinant of the first matrix:  

\begin{equation}
J = bv^3\sin\theta.
\label{eqn:JLIneaire}
\end{equation}

Substituting this expression into equation (\ref{eqn:pxv3}) yields the probability density in the new coordinate system:  

\begin{equation}
p(\theta,\phi,\alpha,v,b,\tau_b) \propto\  bv^3 e^{-v^2/v_T^2}\sin\theta,
\label{eqn:linear}
\end{equation}

\noindent
where $v_T = \sqrt{2kT/m}$ is the thermal velocity, with $m$ denoting the reduced mass. Because there is no interaction with the neutral emitter, the Boltzmann factor accounts only for the kinetic energy.  

This probability distribution serves as the foundation for deriving the joint statistical distributions necessary for particle initialization and the injection of new particles during the simulation.  

\subsection{Statistical Distribution at Initialization of the Simulation}

The purpose of the initialization procedure is to generate a set of perturbers that follow the joint statistical distribution of coordinates described in equation (\ref{eqn:linear}), while ensuring their presence within the simulation volume, which is a sphere of radius $R$.  

Because  
\begin{equation}
{\bf r}(t = 0) = {\bf b} - {\bf v}\tau_b,
\label{eqn:rlin0}
\end{equation}

\noindent the constraint that particles reside inside the sphere implies  

\begin{equation}
|{\bf b} - {\bf v}\tau_b|^2 = b^2 + v^2\tau_b^2 \le R^2,
\end{equation}

\noindent
or, equivalently,
\begin{equation}
|\tau_b| \le t_{\rm max},
\label{eqn:t0tmax}
\end{equation}

\noindent
where  
\begin{equation}
t_{\rm max} \equiv \frac{1}{v}\sqrt{R^2 - b^2}
\label{eqn:tmax}
\end{equation}

\noindent
represents half the particle's crossing time. This constraint on $\tau_b$ introduces statistical dependence among the random variables $(v, b, \tau_b)$.  

By combining the joint statistical distribution (equation \ref{eqn:linear}) with the constraint above, the probability distributions for each coordinate can be derived. The distributions for the three angular coordinates are straightforward:

\begin{eqnarray}
p(\theta) &=& \frac{1}{2}\sin\theta \nonumber\\
p(\phi) &=& \frac{1}{2\pi} \nonumber \\
p(\alpha) &=& \frac{1}{2\pi}
\label{eqn:pangle}
\end{eqnarray}

\noindent
which are properly normalized. Substituting these distributions back into equation (\ref{eqn:linear}) gives the joint probability distribution for the remaining three coordinates, which factorizes as  
\begin{equation}
p(v,b,\tau_b) = p(v)p(b|v)p(\tau_b|b,v)  \propto bv^3 e^{-v^2/v_T^2}.
\label{eqn:linear4}
\end{equation}

\noindent
Because this expression has no explicit dependence on $\tau_b$, the distribution of $\tau_b$ (given $b$ and $v$) is uniform within its allowed range:
\begin{equation}
p(\tau_b|b,v) = \frac{1}{2t_{\rm max}}, \ \ \ \ \ \ {\rm if}\ |\tau_b|\le t_{\rm max}.
\end{equation}

\noindent
Substituting this result into equation (\ref{eqn:linear4}) yields the joint distribution of $v$ and $b$:  
\begin{eqnarray}
p(v)p(b|v)  &\propto& bv^3 e^{-v^2/v_T^2} \times t_{\rm max}  \nonumber \\
&\propto& b\sqrt{R^2 - b^2}\ v^2 e^{-v^2/v_T^2}.
\label{eqn:linear5}
\end{eqnarray}

\noindent
From this, the normalized probability distribution of $b$ (given $v$) is  
\begin{equation}
p(b|v) = \frac{3}{R^3} b\sqrt{R^2 - b^2} \equiv p(b),
\label{eqn:pb}
\end{equation}

\noindent
indicating that $b$ and $v$ are statistically independent. Finally, substituting this result into equation (\ref{eqn:linear5}) and normalizing again gives the Maxwellian distribution of velocity:  

\begin{equation}
p(v)  
=\sqrt{2\over \pi} \frac{1}{v_T^3} v^2 e^{-v^2/v_T^2}.
\label{eqn:pvmaxw}
\end{equation}

The particle generation process at initialization follows these steps. For each particle: (1) the angles $\phi$ and $\alpha$ are drawn from a uniform distribution between 0 and $2\pi$, (2) the angle $\theta$, the impact parameter $b$, and the velocity $v$ are sampled from the distributions given by equations (\ref{eqn:pangle}), (\ref{eqn:pb}), and (\ref{eqn:pvmaxw}), respectively, and (3) the coordinate $\tau_b$ is drawn from a uniform distribution over $\pm t_{\rm max}$, where $t_{\rm max}$ is determined by $b$ and $v$ (equation \ref{eqn:tmax}).  

Once these coordinates have been generated for a set of $N_p$ particles of each type $p$, the simulation begins. At each time step $t_k \equiv k\Delta_t$, 
where \( k \) is an integer and \( \Delta_t \) is the time step,
a particle with coordinates $(\theta,\phi,\alpha,v,b, \tau_b)$ is moved according to  

\begin{eqnarray}
{\bf r}(t_k) &=& b(\cos \alpha {\bf v_1} + \sin \alpha {\bf v_2})
 + v{\bf v_3}(t_k - \tau_b) \nonumber \\
&=& {\bf r}(t_{k-1}) + v{\bf v_3}\Delta_t.
\label{eqn:rti}
\end{eqnarray}

As the simulation progresses, some particles will exit the simulation sphere and must be replaced with new incoming particles of the same type.  

\subsection{Incoming Particles Generating Process}

Historically, finding a reliable procedure for generating incoming particles has been challenging. The initial approach involved creating a new particle with the same $b$ and $v$ values as the outgoing particle, which led to nonphysical behaviors. \citet{Gigosos96} proposed an alternative method based on discrete bins, ensuring that the system’s statistical properties remained constant over time. However, this approach imposed a rigid distribution for $b$ and $v$, eliminating random fluctuations. Moreover, it was only applicable when these coordinates were statistically independent— a condition satisfied for straight-line trajectories but, as will be shown, not for hyperbolic ones.  

\citet{Cho2022} introduced a more general procedure that accounts for statistical dependence between $b$ and $v$ while allowing random fluctuations in both coordinates. The statistical distribution of the incoming particles' $b$ and $v$ coordinates must match that of the outgoing particles. However, particles with longer crossing times through the sphere (i.e., $2t_{\rm max}$) leave less frequently. Consequently, the distribution $p_{\rm in}(v,b)$ of incoming particles must be proportional to the distribution $p(v,b)$ within the sphere (equation \ref{eqn:linear5}), divided by the crossing time:
\begin{equation}
p_{\rm in}(v,b) = p_{\rm in}(v)p_{\rm in}(b|v) \propto p(v)p(b|v) \times \frac{1}{t_{\rm max}},
\end{equation}

\noindent
which yields
\begin{eqnarray}
p_{\rm in}(b|v) &=& \frac{2}{R^2}b \ \ \equiv \ \ p_{\rm in}(b)\nonumber \\
p_{\rm in}(v)  &=& \frac{2}{v_T^4} v^3 e^{-v^2/v_T^2}
\label{eqn:pbvgomez}
\end{eqnarray}

\noindent
after normalization; $b$ and $v$ remain independent variables in the reinjection process for straight-line trajectories, though the reasoning extends to statistically dependent cases.  

When a particle of type $p$ exits the simulation volume at time $t_k$, a new incoming particle of the same type is generated. The angles $\theta, \phi, \alpha$ and the coordinates $b$ and $v$ are drawn from their respective distributions (equations \ref{eqn:pangle} and \ref{eqn:pbvgomez}). 
The coordinate $\tau_b$ is then determined by requiring that the incoming particle be located on the surface of the simulation volume, or within a thin shell adjacent to it, at time $t_k$.

\section{Hyperbolic Trajectory: Generalization of the Standard Method  \label{sec:hyperb}}

The equations used to generate particles with hyperbolic trajectories are derived in this section for the TMD model. We draw inspiration from the proven method for straight-line trajectories, for which the six new coordinates serve two roles: they constitute a coordinate system in phase space (at each time $t$), and they parameterize all possible trajectories as a function of time. The first step is therefore to uniquely define a hyperbolic trajectory with six parameters, and show they adequately cover the entire phase space volume relevant for the simulation. The jacobian of the coordinate transformation from the Cartesian coordinates $({\bf r,\ v})$ in phase space to the set of new coordinates is determined, along with the relevant statistical distributions used to generate particles at initialization and incoming particles.

\subsection{Geometry of the Hyperbola}

A hyperbola centered at the origin with its principal axes aligned along the $x$- and $y$-coordinates is given by:
\begin{equation}
{\bf r} = a \cosh(u) \ {\bf x}  + b \sinh(u) \ {\bf y},
\label{eqn:h3} 
\end{equation}

\noindent
where $u$ is the hyperbolic eccentric anomaly, and $a$ and $b$ are the semi-major and semi-minor axes, respectively. In gravitational trajectory studies, the convention $a < 0$ is often used to unify elliptical and hyperbolic orbit equations. However, in this work, we adopt $a > 0$ to avoid confusion with the signs that appear in subsequent equations.

Equation (\ref{eqn:h3}) must be modified for simulation purposes by shifting the trajectory so that the source, originally at a focus, is moved to the origin:
\begin{eqnarray}
{\bf r}&= & a \cosh(u) \ {\bf x} + b \sinh(u)\ {\bf y} +\lambda \sqrt{a^2+b^2}\ {\bf x}\nonumber \\
&=&a\bigg((\cosh(u)+\lambda e){\bf x} + \sqrt{e^2-1} \sinh(u ){\bf y}\bigg),
\label{eqn:h4} 
\end{eqnarray}

\noindent
where $\lambda$ represents the sign of the electric force ($-1$ for attraction and $+1$ for repulsion), and $e$ is the eccentricity:
\begin{equation}
e = {\sqrt{a^2+b^2} \over a} > 1.
\label{eqn:e}
\end{equation}

\noindent
Given that $b/a = \sqrt{e^2-1}$, the radial distance from the origin follows as:
\begin{equation}
r =  a(e\cosh(u)+\lambda),
\label{eqn:r2} 
\end{equation}

\noindent
which is always positive because $e\cosh(u) > 1$. The minimum distance, $\sigma$, occurs at $u=0$, where $\cosh(u)$ is minimized:
\begin{equation}
\sigma= a(e+\lambda).
\label{eqn:sigma}
\end{equation}

\noindent
A key characteristic of a hyperbolic trajectory is its asymptote and associated impact parameter. As $u \to \pm\infty$, the trajectory approaches two straight lines with a minimum distance to the origin of $b$, meaning the asymptotic impact parameter is precisely the hyperbola’s $b$ parameter.

To obtain the most general trajectory, three rotations are applied. First, the trajectory in equation (\ref{eqn:h4}) undergoes a rotation by an angle $\beta$ in the $x$-$y$ plane. Then, the normal ${\bf z}$ of the orbital plane is rotated to align with:
\begin{eqnarray}
{\bf w_3}&=&(\cos\phi \sin\theta, \sin\phi \sin\theta, \cos\theta)^\text{T},
\label{eqn:w3} 
\end{eqnarray}

\noindent
where $0 \leq \phi < 2\pi$ and $0 \leq \theta \leq \pi$. After applying these transformations, the general trajectory becomes:
\begin{equation}
{\bf r} = a \bigg((\cosh(u) +\lambda e) {\bf w_1}+ \sqrt{e^2-1}\sinh(u){\bf w_2}\bigg),
\label{eqn:rr} 
\end{equation}

\noindent
where:
\begin{eqnarray}
{\bf w_1} &=& \cos\beta {\bf v_1} + \sin\beta {\bf v_2},\nonumber\\
{\bf w_2} &=& -\sin\beta {\bf v_1} + \cos\beta {\bf v_2},
\label{eqn:w12} 
\end{eqnarray}
\noindent
and:
\begin{eqnarray}
{\bf v_1}&=&(\cos\phi \cos\theta, \sin\phi \cos\theta, -\sin\theta)^\text{T},\nonumber \\
{\bf v_2}&=&(-\sin\phi , \cos\phi, 0)^\text{T}.
\label{eqn:v123} 
\end{eqnarray}

\noindent
The vectors ${\bf w_{1,2,3}}$ form an orthonormal basis that coincides with the $x$, $y$, and $z$ axes when $\theta = \phi = \beta = 0$.

\subsection{Dynamics of the Hyperbolic Trajectory \label{subsec:dynamic}}

The equations in the previous section described the trajectory’s geometry without considering its dynamics. We now incorporate energy conservation to determine how the implicit parameter $u$ depends on time $t$.  

In a coordinate system where the emitter (charge $q_e$) is stationary, the total energy of a perturbing particle with reduced mass $m$, charge $q_p$, and relative velocity $v$ is the sum of kinetic and potential energies:
\begin{eqnarray}
E &=& {mv^2\over 2} + {q_eq_p\over r} \nonumber \\
 &=& m\bigg({v^2\over 2} + \lambda {\kappa\over r} \bigg),
\label{eqn:E0} 
\end{eqnarray}

\noindent
where $\lambda$ represents the sign of $q_eq_p$, determining whether the interaction is attractive or repulsive. The interaction intensity parameter $\kappa$ is always positive and defined as:

\begin{equation}
\kappa \equiv {|q_eq_p|\over m}.
\label{eqn:k} 
\end{equation}

\noindent
Energy is of course conserved and takes the same value (as a function of $a$) regardless of whether the force is attractive or repulsive:
\begin{equation}
E = {m\kappa\over 2a}.
\label{eqn:E} 
\end{equation}

\noindent
Combining equations (\ref{eqn:E0}) and (\ref{eqn:E}) gives the relation between distance $r$ and velocity $v$:
\begin{equation}
\kappa\bigg({1\over a} - \lambda {2\over r}\bigg) = v^2,
\label{eqn:E3} 
\end{equation}

\noindent
from which we obtain the velocity at infinity:
\begin{equation}
v_\infty^2 = \lim_{r\rightarrow \infty} \kappa\bigg({1\over a} - \lambda {2\over r}\bigg) = {\kappa\over a}.
\label{eqn:E4} 
\end{equation}

\noindent
The velocity vector ${\bf v}$ is given by the time derivative of equation (\ref{eqn:rr}):
\begin{equation}
{\bf v} \equiv {d{\bf r}\over dt} = a{du\over dt} (\sinh(u){\bf w_1} + \sqrt{e^2-1}\cosh(u){\bf w_2}),
\label{eqn:v} 
\end{equation}

\noindent
where the standard derivatives of hyperbolic functions have been used. The time derivative of $u$ follows from equations (\ref{eqn:r2}), (\ref{eqn:E3}), and (\ref{eqn:v}):

\begin{equation}
{du\over dt} = \sqrt{\kappa\over a^3}{1\over (e\cosh(u)+\lambda)},
\label{eqn:FT2} 
\end{equation}

\noindent
where we take the positive square root to ensure that $u$ increases with time. Integrating this equation gives the time dependence of $u$:
\begin{equation}
e\sinh(u) +\lambda u = \sqrt{\kappa\over a^3}(t-\tau_b),
\label{eqn:FT_final}
\end{equation}

\noindent
where $\tau_b$ is an integration constant, identified as the time of closest approach (when $u=0$). Finally, substituting this result into equation (\ref{eqn:v}) yields an alternative expression for the velocity vector:
\begin{equation}
{\bf v} = \sqrt{\kappa\over a} {1 \over e\cosh(u)+\lambda}  (\sinh(u){\bf w_1} + \sqrt{e^2-1}\cosh(u){\bf w_2}).
\label{eqn:v_final} 
\end{equation}

\subsection{Jacobian of the Hyperbolic Impact Coordinate Transformation \label{sec:determ}}

The next step is to identify the set of phase-space coordinates that simultaneously represent every accessible point in phase space (at any given time \( t \)) and describe every hyperbolic trajectory of a perturber. Several parameters can characterize a hyperbolic trajectory in addition to the three angular parameters: the particle's energy, the velocity at infinity, the eccentricity, the semi-major and the semi-minor axes, among others. We impose the constraint that the particle generation method must converge to the case of straight-line trajectories when the interaction intensity parameter \( \kappa \) vanishes.  

The most natural generalization of the velocity coordinate \( v \) of a straight-line path is the velocity at infinity, \( v_\infty \). Two possible choices exist for the coordinate corresponding to the impact parameter in straight-line trajectories: the minimum distance to the origin, \( \sigma \), and the asymptotic impact parameter, \( b \). It turns out that the latter naturally leads to a joint statistical distribution that converges to the case of straight-line trajectories. The sixth coordinate adopted is the time of closest approach, \( \tau_b \), which corresponds to its equivalent in the context of straight-line trajectories.  

It can be shown that the six proposed parameters, \( {\bf x} \equiv (\theta, \phi, \alpha, v_\infty, b, \tau_b) \), provide a valid coordinate system for the hyperbolic trajectory because each point \(({\bf r, v})\) in phase space corresponds to a unique solution \( {\bf x} \) at any constant value of \( t \). The only exception is the set of points with negative energy. 
However, these points were not considered as admissible perturber positions within the numerical simulation.

As discussed for straight-line trajectories (equation \ref{eqn:pxv3}), the probability density of the particle's position in the new coordinates is given by the product of a Boltzmann factor and the jacobian of the coordinate transformation. For hyperbolic trajectories, the jacobian matrix \( {\bf J} \) is expressed as  
\begin{equation}
{\bf J} \equiv \left (\begin{matrix}
{\partial {\bf r^\text{T}} \over \partial v_\infty} & {\partial {\bf v^\text{T}} \over \partial v_\infty} \\
{\partial {\bf r^\text{T}} \over \partial b} &{\partial {\bf v^\text{T}} \over \partial b}  \\
{\partial {\bf r^\text{T}} \over \partial \tau_b} &{\partial {\bf v^\text{T}} \over \partial \tau_b}  \\
{\partial {\bf r^\text{T}} \over \partial \theta} &{\partial {\bf v^\text{T}} \over \partial \theta}  \\
{\partial {\bf r^\text{T}} \over \partial \phi} &{\partial {\bf v^\text{T}} \over \partial \phi}  \\
{\partial {\bf r^\text{T}} \over \partial \beta} &{\partial {\bf v^\text{T}} \over \partial \beta}  \end{matrix}
\right )
\label{eqn:Jmatrix}
\end{equation}
where, following the same notation as in equation (\ref{eqn:Jstraighline}), the elements of the matrix are written as transposed 3D-vectors.
    
After systematic though tedious algebra (see Appendix \ref{sec:determ2}), evaluating the determinant of ${\bf J}$ leads to the jacobian:
\begin{equation} 
J = bv^3_\infty \sin\theta,
\end{equation}

\noindent
which, when substituted into equation (\ref{eqn:pxv3}), yields the probability distribution for the particle’s position in the new coordinate system:
\begin{equation}
p(\theta,\phi,\alpha,v_\infty,b,\tau_b) \propto\  b v_\infty^3 e^{-v_\infty^2/v_T^2} \sin\theta.
\label{eqn:bbvts0}
\end{equation}

\noindent
Surprisingly, this probability distribution retains the same form for both straight-line trajectories at constant velocity \( v \) and hyperbolic trajectories. This holds as long as a correspondence is made between \( v \) and \( v_\infty \), and between the impact parameter \( b \) in both cases. This expression serves as the foundation for deriving the joint statistical distributions needed for both the initialization of particle generation and the injection of new particles during the simulation.

\subsection{The Probability Distribution at Initialization}

The goal of this section is to determine the joint statistical distribution of the six coordinates ($\theta,\phi,\alpha, v_\infty, b, \tau_b$) required for generating particles at the initialization ($t = 0$) of the simulation. The statistical distribution (equation \ref{eqn:bbvts0}) is subject to the constraint that particles must be inside the sphere of radius $R$. Additionally, at each step, we allow the interaction intensity parameter $\kappa$ to converge to 0, ensuring that in the absence of interaction (i.e., a neutral emitter), the solution corresponds to a straight-line trajectory. It is important to note that the minimum distance $\sigma$ of the hyperbolic trajectory from the origin is not identical to the impact parameter $b$.  

Given the isotropy of the problem, as inferred from the angular dependence of the statistical distribution (equation \ref{eqn:bbvts0}), the generation of the three angles follows equation (\ref{eqn:pangle}), derived for straight-line trajectories. However, the generation of the remaining three coordinates deviates from this simple case.  

The minimum distance to the origin is expressed in terms of the variables $v_\infty, b$, and $\tau_b$:  
\begin{equation}
\sigma = a(e+\lambda) =\sqrt{a^2+b^2} + \lambda a,
\end{equation}

\noindent
where $e$ is the eccentricity (equation \ref{eqn:e}), $\lambda = \pm 1$, and $a = {\kappa/ v_\infty^2}$. The constraint requiring particles to be within the simulation sphere implies that  
\begin{equation}
\sigma = \sqrt{a^2+b^2} + \lambda a \le R,
\end{equation}

\noindent
which leads to the condition  
\begin{equation}
R \ge 2a\ \max (0, \lambda).
\label{eqn:bRcontraint}
\end{equation}

\noindent
This result has distinct implications depending on the sign of $\lambda$. When $\lambda = -1$, the condition reduces to the trivial result $R \ge 0$. However, when $\lambda = +1$, it imposes a constraint on $v_\infty$, determining the minimum velocity required for a particle to enter the sphere:  
\begin{equation}
v_\infty \ge \sqrt{2\kappa\over R}.
\end{equation}

\noindent
Thus, low-energy particles cannot enter the sphere due to the repulsive interaction. Combining these results for both values of $\lambda$, the minimum velocity (at infinity) required for a particle to enter the sphere is  
\begin{equation}
v_{\rm min} \equiv {\rm Max}\bigg( 0,  \lambda \sqrt{2\kappa\over R}\bigg).
\label{eqn:vmin}
\end{equation}

Following the same approach used for straight-line trajectories, the distribution of the three dependent variables is factored into a product of distributions:  
\begin{equation}
p(v_\infty,b,\tau_b) = p(v_\infty)p(b|v_\infty)p(\tau_b|b, v_\infty) \propto
e^{-v_\infty^2/ v_T^2}  v^3_\infty b.
\label{eqn:bayes2}
\end{equation}

\noindent
The constraint that particles must be present in the sphere (at initialization) is expressed by an inequality involving the implicit variable $u$ of the hyperbolic trajectory, starting from equation (\ref{eqn:r2}):  
\begin{eqnarray}
r(u) = a(e\cosh(u) + \lambda) \le R 
\rightarrow |u| &\le& \cosh^{-1} \bigg({R-\lambda a\over \sqrt{a^2+b^2}}\bigg)
 \equiv u_{\rm max}.
\end{eqnarray}

\noindent
This result is converted into a constraint on the time of closest approach by using the relation between $u$ and $\tau_b$ (equation \ref{eqn:FT_final}) for the particular case $t=0$:  
\begin{equation}
\tau_b = t -\sqrt{a^3\over \kappa}(e\sinh(u) +\lambda u) \rightarrow {v\over a}\tau_b = -(e\sinh(u) +\lambda u).
\label{eqn:b0_2}
\end{equation}

\noindent
The constraint $|u| \le u_{\rm max}$ implies that $|\tau_b| \le t_{\rm max}$, with

\begin{equation}
t_{\rm max} = {1\over v}\bigg(\sqrt{(R-\lambda a)^2 -({a^2+b^2})} + 
\lambda a\cosh^{-1}\bigg({R-\lambda a\over \sqrt{a^2+b^2}}\bigg)  \bigg),
\label{eqn:tmaxH}
\end{equation}

\noindent
which is a function of $v_\infty$ and $b$. For a neutral emitter, the interaction intensity parameter $\kappa$ (and therefore $a$) vanishes, reducing the half-crossing time to the solution for a straight-line path:  
\begin{equation}
t_{\rm max} = {1\over v}\sqrt{R^2 - b^2} \nonumber.
\end{equation}

Because $p(v_\infty,b,\tau_b)$ has no explicit dependence on $\tau_b$, the probability distribution $p(\tau_b|b, v_\infty)$ is uniform within its domain:  
\begin{equation}
p(\tau_b|b, v_\infty) = {1\over 2t_{\rm max}} \quad \text{with} \ |\tau_b| < t_{\rm max}.
\end{equation}  

\noindent  
Substituting this result back into equation (\ref{eqn:bayes2}) gives the joint distribution of \( v_\infty \) and \( b \):  

\begin{eqnarray}  
p(v_\infty)p(b|v_\infty) &\propto& e^{-v_\infty^2/ v_T^2} v^3_\infty b\ \times \ t_{\rm max} \nonumber \\  
&=&  b \bigg(\sqrt{(R-\lambda a)^2 - (a^2+b^2)} + \lambda a \cosh^{-1}\bigg({R-\lambda a\over \sqrt{a^2+b^2}}\bigg) \bigg) e^{-v_\infty^2/ v_T^2} v_\infty^2. \nonumber \\  
&&  
\label{eqn:bpvsigma}  
\end{eqnarray}  

\noindent  
The requirement that the square root remains real imposes a constraint on the impact parameter \( b \) for a particle to enter the sphere:  
\begin{eqnarray}  
(R-\lambda a)^2 - (a^2+b^2) &\ge& 0 \nonumber \\  
\rightarrow b &\le& \sqrt{R(R - 2\lambda a)} \equiv b_{\rm max},  
\label{eqn:bmax}  
\end{eqnarray}  

\noindent  
where \( b_{\rm max} \) depends on \( v_\infty \). The distribution \( p(b|v_\infty) \) can be determined by isolating terms involving \( b \):  
\begin{equation}  
p(b|v_\infty) \propto b \bigg( \sqrt{(R-\lambda a)^2 - (a^2+b^2)} + \lambda a \cosh^{-1}\bigg({R-\lambda a\over \sqrt{a^2+b^2}}\bigg) \bigg). 
\end{equation}  

\noindent  
To normalize this distribution, we compute the normalization constant \( K_B \), given by the following integral over the allowed range of \( b \), separated into two terms, \( I_1 \) and \( I_2 \):  

\begin{eqnarray}  
K_B &\equiv& \int_0^{b_{\rm max}} db\ b \bigg(\sqrt{(R-\lambda a)^2 - (a^2+b^2)} + \lambda a \cosh^{-1}\bigg({R - \lambda a \over \sqrt{a^2+b^2}} \bigg) \nonumber \\  
&=& I_1 + \lambda a I_2,  
\end{eqnarray}  

\noindent  
where both terms have analytical solutions:  
\begin{eqnarray}  
I_1 &=& {1\over 3}\bigg((R - \lambda a)^2 - a^2\bigg)^{3\over 2} \nonumber \\  
I_2 &=&  {1\over 2} \bigg( (R - \lambda a) \sqrt{(R - \lambda a)^2 - a^2} - a^2 \cosh^{-1}\bigg({R - \lambda a \over a}\bigg) \bigg).  
\end{eqnarray}  

\noindent  
Thus, the properly normalized distribution is  
\begin{equation}  
p(b|v_\infty) = K_B^{-1} \times b \bigg( \sqrt{(R-\lambda a)^2 - (a^2+b^2)} +\lambda a\cosh^{-1}\bigg({R-\lambda a\over \sqrt{a^2+b^2}}\bigg) \bigg),  
\label{eqn:bvH32}  
\end{equation}  

\noindent  
which implies that \( b \) and \( v_\infty \) are not statistically independent, in contrast to the case of a straight-line trajectory.  

For a neutral emitter, where there is no electrical interaction (\( \kappa = 0, \ a = 0 \)), the integral simplifies to \( I_1 = {1\over 3}R^3 \) and \( \lambda aI_2 = 0 \). Consequently, the resulting distribution reduces to  
\begin{eqnarray}  
p(b|v) &=& 3{b \over R^3} \sqrt{R^2 - b^2}, \nonumber  
\end{eqnarray}  

\noindent  
which recovers the well-known result for a straight-line trajectory, demonstrating its independence from \( v \).   
Substituting equation (\ref{eqn:bvH32}) into (\ref{eqn:bpvsigma}) and normalizing gives the velocity distribution at infinity:
\begin{eqnarray}  
p(v_\infty) &=& K_V^{-1}\ e^{-v_\infty^2/ v_T^2} v^2 \times \big( I_1(R, \lambda a) + \lambda a I_2(R, \lambda a)\big), \label{eqn:vH32}  
\end{eqnarray}  

\noindent  
where \( a = k/v_\infty^2 \), and the normalization constant \( K_V \) is determined through numerical integration over the valid velocity range:  
\begin{equation}  
K_V \equiv \int_{v_{\rm min}}^\infty p(v_\infty)\,dv_\infty = \int_{v_{\rm min}}^\infty dv\ e^{-v^2/ v_T^2} v^2 \times \Bigg( I_1\bigg(R, \lambda {\kappa\over v^2}) + \lambda {\kappa \over v^2} I_2\bigg(R, \lambda {\kappa\over v^2}\bigg)\Bigg),  
\end{equation}  

\noindent  
where \( v_{\rm min} \) is given by equation (\ref{eqn:vmin}).  

The initialization of the particle generation process follows these steps. For each particle, the angles \( \phi \) and \( \alpha \) are drawn from a uniform distribution between 0 and \( 2\pi \). The angle \( \theta \) and velocity \( v_\infty \) are then sampled from equations (\ref{eqn:pangle}) and (\ref{eqn:vH32}), respectively. Once \( v_\infty \) is known, the impact parameter \( b \) is determined using equation (\ref{eqn:bvH32}). Finally, given \( v_\infty \) and \( b \), the time of closest approach \( \tau_b \) is drawn from a uniform distribution between \( \pm t_{\rm max} \), as defined in equation (\ref{eqn:tmaxH}). The numerical details of this random generation process are discussed in Appendix \ref{app:I}.  

Once the coordinates for a set of \( N_p \) particles of each type \( p \) are generated, the simulation begins. At each time step \( t_k \equiv k\Delta_t \), 
where \( k \) is an integer and \( \Delta_t \) is the time increment, the particle positions are updated according to:  
\begin{eqnarray}  
{\bf r}(t_k) &=& a\bigg((\cosh(u_k) + \lambda e){\bf w_1} + \sqrt{e^2-1}\sinh(u_k) {\bf w_2}\bigg), \label{eqn:rtiH}  
\end{eqnarray}  

\noindent  
where  
\begin{eqnarray}  
a &=& {\kappa\over v_\infty^2}, \nonumber \\  
e &=& {\sqrt{a^2+b^2}\over a}.  
\end{eqnarray}  

\noindent  
The vectors \( {\bf w_1} \) and \( {\bf w_2} \) depend on the three angles \( (\theta, \phi, \alpha) \), while \( u_k \) is the solution to the nonlinear equation:  
\begin{eqnarray}  
e\sinh(u_k) + \lambda u_k &=& \sqrt{\kappa\over a^3}(t_k - \tau_b). \label{eqn:Fi}  
\end{eqnarray}  

\noindent  
Thus, computing the particle's position at time \( t_k \) requires solving for \( u_k \), a procedure detailed in Appendix \ref{sec:AppHyper}.  

\subsection{The Probability Distribution of Incoming Particles}

\noindent  
The probability distribution of incoming particles is obtained by extending the method of \citet{Cho2022} to hyperbolic trajectories. The distribution \( p_{\rm in}(v_\infty, b) \) of incoming particles must be proportional to the distribution \( p(v_\infty, b) \) of particles inside the sphere (equation \ref{eqn:bpvsigma}), divided by the crossing time:
\begin{equation}  
p_{\rm in}(v_\infty,b) = p_{\rm in}(v_\infty)p_{\rm in}(b|v_\infty) \propto p(v_\infty)p(b|v_\infty) \times {1\over t_{\rm max}}.  
\end{equation}  

\noindent  
The result retains the same form as in the case of straight-line trajectories:
\begin{equation}  
p_{\rm in}(v_\infty)p_{\rm in}(b|v_\infty) \propto e^{-{v_\infty^2/v_T^2}} v_\infty^3 b,  
\label{eqn:phb}  
\end{equation}  

\noindent  
which implies that  
\begin{equation}  
p_{\rm in}(b|v_\infty) \propto b.  
\end{equation}  

\noindent  
The normalization constant is obtained by solving the integral  

\begin{equation}  
\int_0^{b_{\rm max}} b \ db = {1\over 2}b_{\rm max}^2,  
\end{equation}  

\noindent  
where \( b_{\rm max} \) is defined in equation (\ref{eqn:bmax}). The normalized distribution is then  
\begin{equation}  
p_{\rm in}(b|v_\infty) = {2b\over b^2_{\rm max}} = {2\over R}{b \over (R - 2\lambda a)},  
\label{eqn:pinb}  
\end{equation}  

\noindent  
with \( a = \kappa/v_\infty^2 \). Unlike in the case of straight-line trajectories, \( b \) and \( v_\infty \) are not independent variables.  

Substituting this result into equation (\ref{eqn:phb}) yields the (unnormalized) velocity distribution at infinity for particles entering the sphere:  
\begin{equation}  
p_{\rm in}(v_\infty) \propto e^{-{v_\infty^2/v_T^2}} v_\infty \bigg(v_\infty^2 - 2\lambda {\kappa\over R}\bigg),  
\end{equation}  

\noindent  
which is valid for \( v_\infty > v_{\rm min} \), as defined in equation (\ref{eqn:vmin}). Normalizing this expression gives the velocity distributions for incoming particles:  
\begin{eqnarray}  
\lambda = -1,\ \ v_\infty \ge 0: \quad p_{\rm in}(v_\infty) &=& {2\over v_T^4}  
\ e^{-{(v_\infty^2/v_T^2})} v_\infty^3 {1 + 2 {\kappa\over Rv_\infty^2} \over  1+2 {\kappa\over Rv_T^2}}, \nonumber \\  
\lambda = +1,\ \ v_\infty \ge v_{\rm min}: \quad p_{\rm in}(v_\infty) &=&  
{2\over v_T^4 }\ e^{-{(v_\infty^2-v_{\rm min}^2)/v_T^2}} \times v_\infty \bigg(v_\infty^2 - v_{\rm min}^2\bigg),  
\label{eqn:pinv}  
\end{eqnarray}  

\noindent  
where \( v_{\rm min} = \sqrt{2\kappa\over R} \) for \( \lambda = +1 \), and \( v_T \) is the thermal velocity. The result for straight-line trajectories is recovered when \( \kappa = 0 \).  

When a particle of type \( p \) exits the simulation volume at time \( t_k \), a new incoming particle must be generated. The angles \( \theta, \phi, \alpha \) and the coordinates \( b \) and \( v_\infty \) are drawn from their respective distributions (equations \ref{eqn:pangle}, \ref{eqn:pinb}, and \ref{eqn:pinv}). The numerical details of this random generation process are discussed in Appendix \ref{app:III}.  
The remaining coordinate \( \tau_b \) is defined so that, at time $t_k$, the particle is located precisely on the surface of the simulation volume.

\section{Results}\label{sec:comp}

\subsection{Statistical Distribution of the Coordinates}\label{subsec:dist}

The first validation tests of the simulation consist of examining the statistical distributions of the perturber coordinates—ions and electrons—and of the resulting electric-field intensities, both at initialization and as a function of time.

\begin{figure}
\centering
\includegraphics[width=1\linewidth]
{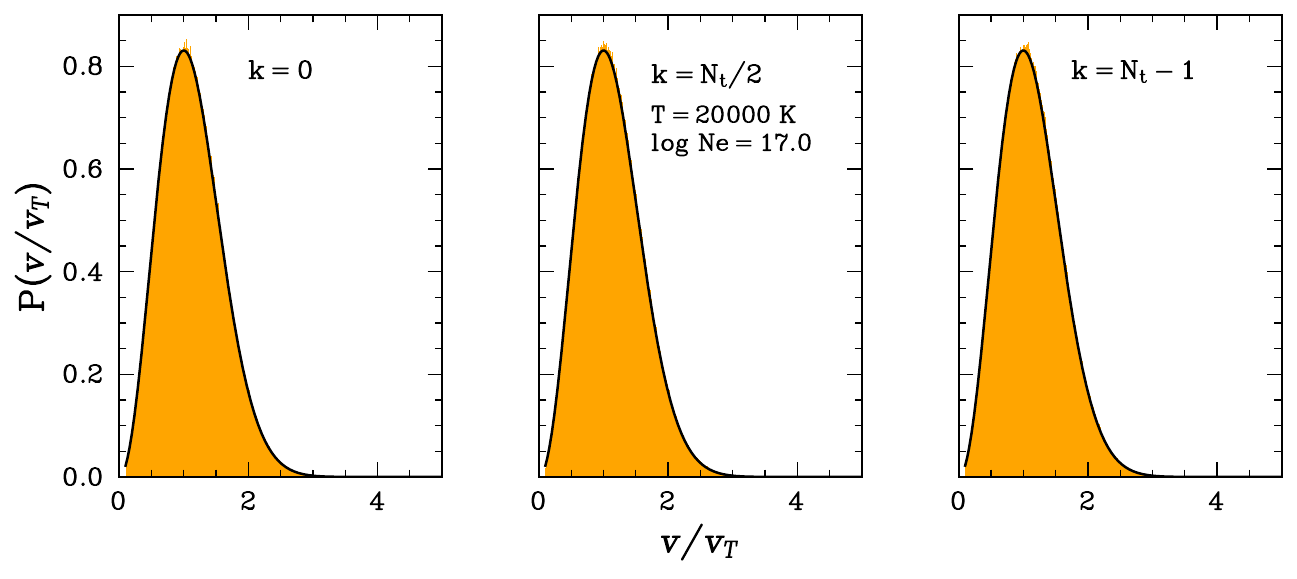}
\includegraphics[width=1\linewidth]
{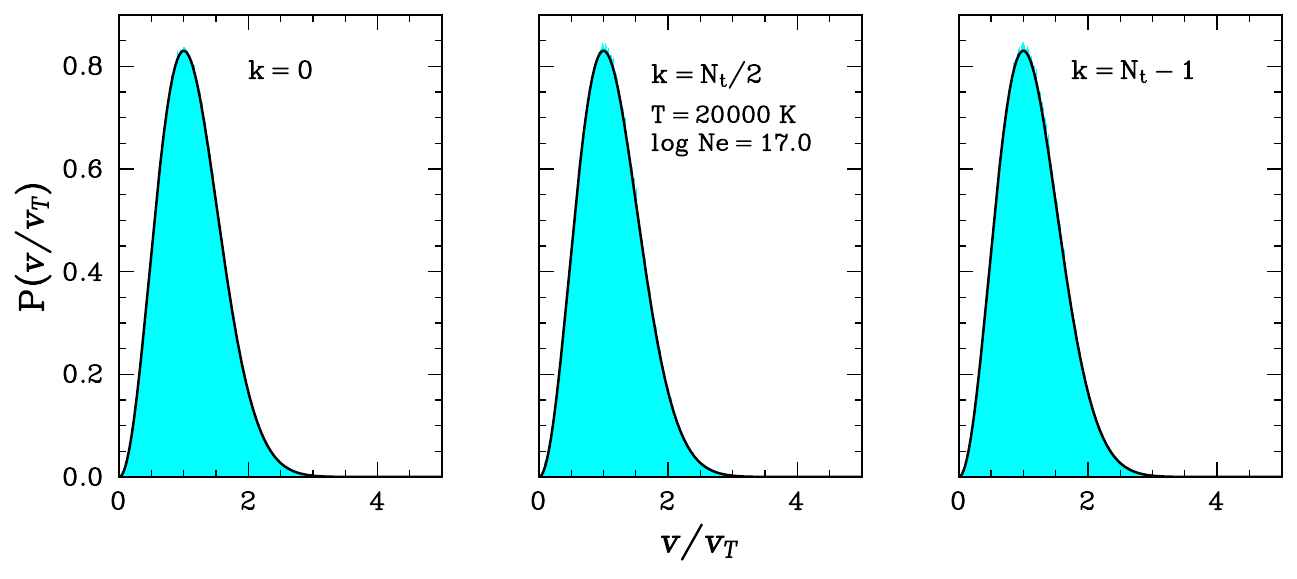}
\caption[Statistical distribution of the velocity at infinity of ions
  and electrons]{ Statistical distribution of the velocity at infinity
  of ions (top) and electrons (bottom) for a temperature of 20,000 K
  and an electronic density of $10^{17}$ cm$^{-3}$ at three time-step
  indices $k=t_k/\Delta_t$ of 0, $N_t/2$, and $N_t-1$. The velocity
  $v$ is normalized by the thermal velocity $v_T$. The theoretical
  distributions are shown by the black lines, while the generated
  distributions for ions and electrons are represented by the orange
  and cyan histograms, respectively.  }
\label{fig:vinfpaperIII}
\end{figure}

\begin{figure}
\centering
\includegraphics[width=1\linewidth]
{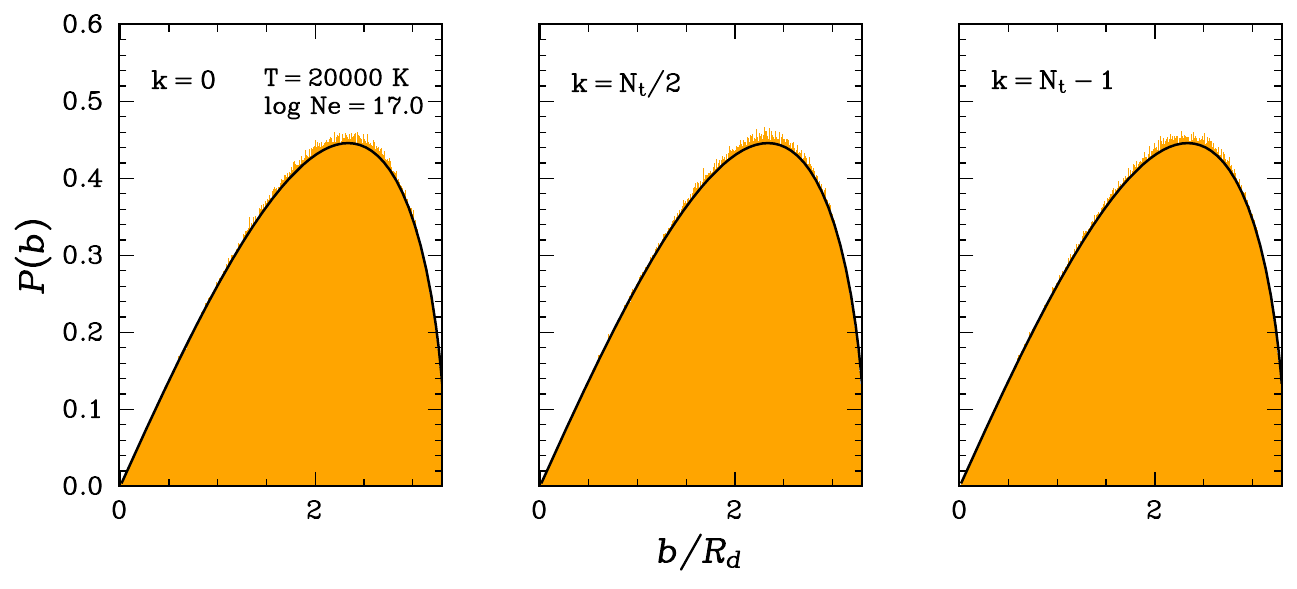}
\includegraphics[width=1\linewidth]
{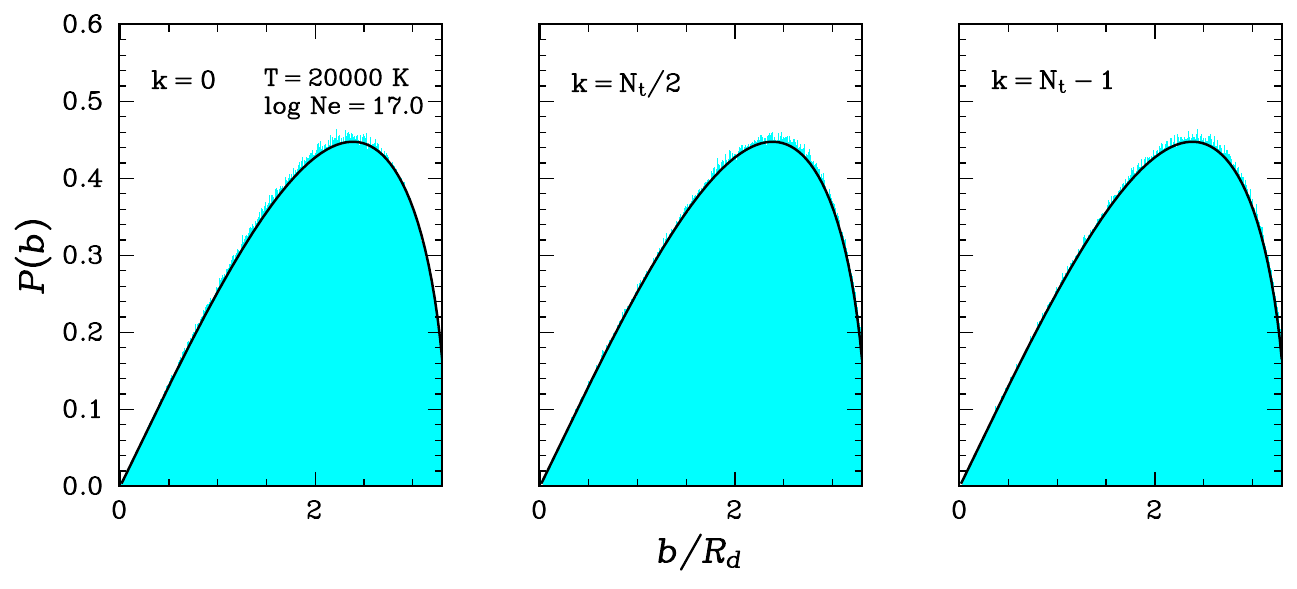}
\caption[Statistical distribution of the impact parameter for ions and
  electrons]{Statistical distribution of the impact parameter for ions
  (top) and electrons (bottom) for a temperature of 20,000 K and an
  electronic density of $10^{17}$ cm$^{-3}$ at three time-step indices
  $k=t_k/\Delta_t$ of 0, $N_t/2$, and $N_t-1$. The impact parameter
  $b$ is normalized by the Debye radius $R_d$.  The theoretical ion
  and electron functions are shown in black, whereas the corresponding
  generated functions are represented by the orange and cyan
  histograms, respectively.}
\label{fig:BstatpaperIII}
\end{figure}

Figure \ref{fig:vinfpaperIII} illustrates the distribution of velocities at infinity, averaged over all simulation volumes, at initialization, at mid-time, and at the end of the simulation, shown separately for ions and electrons.
The distributions follow the Maxwell distribution, except in the case of ions, where particles with insufficient energy to enter the simulation volume are excluded because of the repulsive interaction with the emitting ion, as defined by equation (\ref{eqn:vmin}). A similarly good agreement between the theoretical and numerical distributions is obtained for the impact parameter $b$, as shown in Figure \ref{fig:BstatpaperIII}.

\begin{figure}
\centering
\includegraphics[width=1\linewidth]
{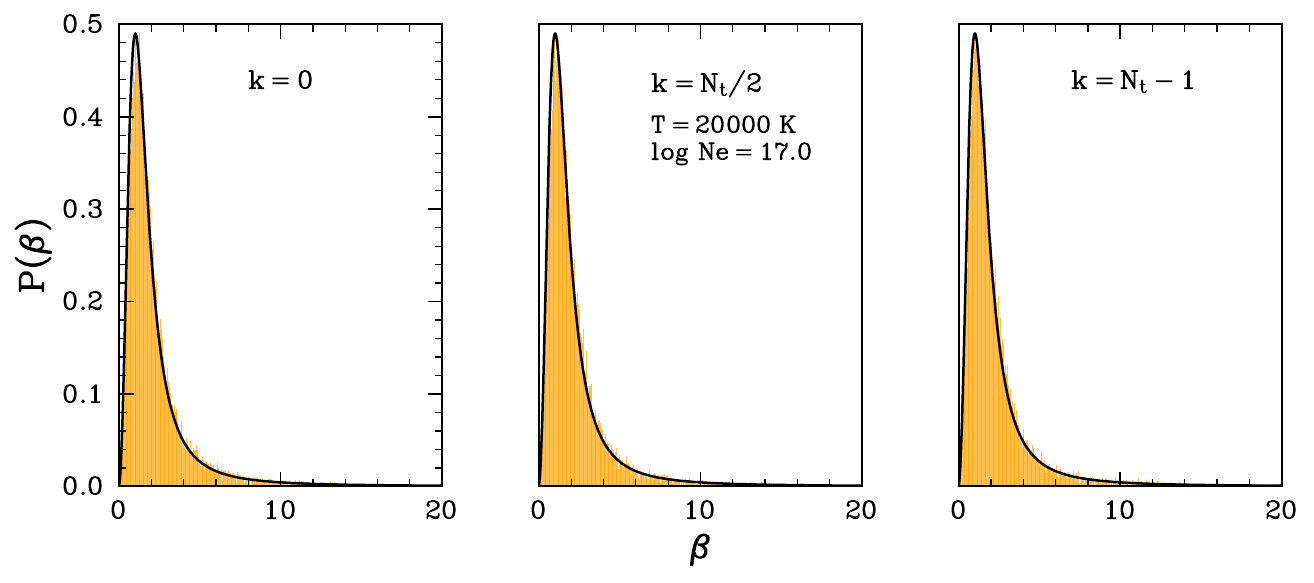}
\includegraphics[width=1\linewidth]
{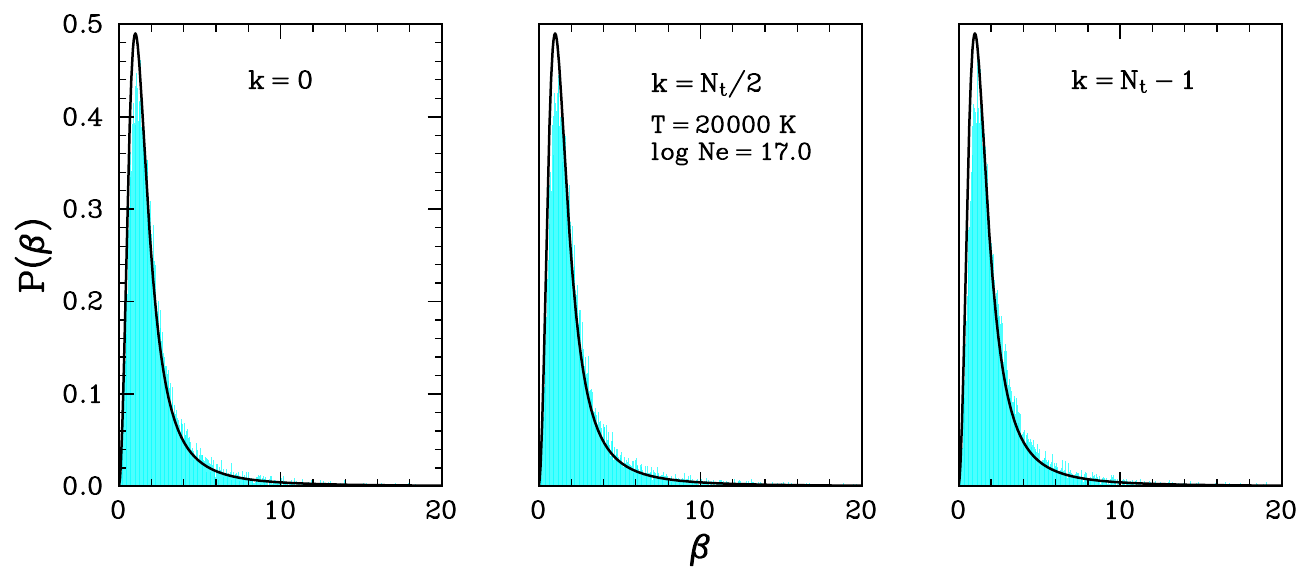}
\caption[Statistical distribution of the electric field intensity for ions and electrons]{Statistical distribution of the electric field intensity for ions (top) and electrons (bottom)
for a temperature of 20,000 K and   an electronic density of $10^{17}$ cm$^{-3}$ at three time-step   indices $k=t_k/\Delta_t$ of 0, $N_t/2$, and $N_t-1$. $\beta$ is the intensity of the electric field normalized by the (ionic) Holtzmark field.  The distributions from \citet{Hooper68} are shown in black, while the generated   distributions for ions and electrons are represented by the orange and cyan histograms, respectively.}
\label{fig:EFieldpaperIII}
\end{figure}

The distribution of the electric-field intensity produced by the ions is consistent with the Hooper distribution, derived for the environment of an ion with charge $Z = +1$ \citep{Hooper68}, as illustrated in Figure
\ref{fig:EFieldpaperIII} for three characteristic times. 
The electron distribution exhibits a more pronounced contribution in the wings, resulting in a reduced peak intensity. This behavior arises from the attractive interaction, which favors the presence of electrons near the emitting ion.

A diagnostic that is sensitive both to the random generation of the particle coordinates 
and to the history of particle motion is the pair-correlation function $g(r)$, also known as the radial distribution function. It quantifies the probability of finding a charged perturber at a distance $r$ from the emitting radiator relative to the case of a uniform distribution. In this way, $g(r)$ measures the local modification of the particle density around the emitter induced by electrostatic interactions.

\begin{figure}[t]
\centering
\includegraphics[width=1\linewidth]
{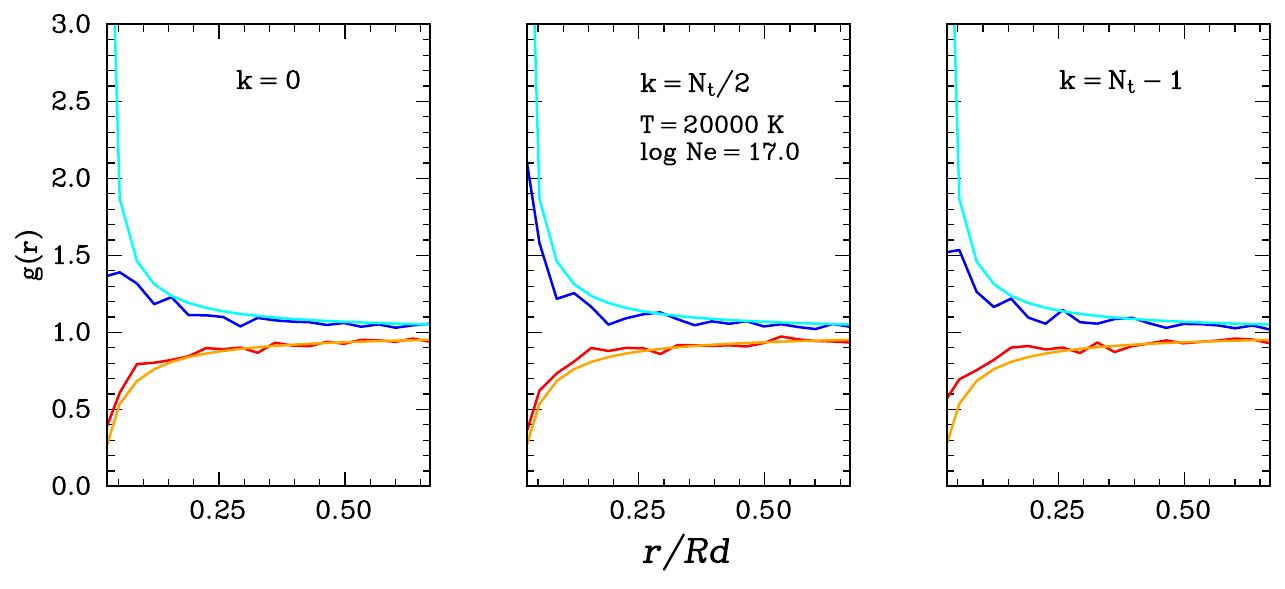}
\caption[Statistical distribution of $g(r)$ for ions and electrons]{ Statistical distribution of $g(r)$ for ions and electrons 
for a temperature of 20,000 K and
  an electronic density of $10^{17}$ cm$^{-3}$ at three time-step
  indices $k=t_k/\Delta_t$ of 0, $N_t/2$, and $N_t-1$. The distance $r$ is normalized by the Debye radius $R_d$. 
  The theoretical ion and electron functions are shown in orange and cyan, respectively, whereas the corresponding generated functions are depicted in red and blue.}
\label{fig:grpaperIII}
\end{figure}

Assuming an interaction potential $U(r)$ between the perturbers of a given type and the emitter,
$g(r)$ is defined as
\begin{equation}
g(r) = e^{-U(r)/kT}.
\end{equation}
Because the perturbers' trajectories are generated assuming the full Coulomb interaction with the emitter rather than a screened Debye potential,
\begin{equation}
g(r) = {\rm exp}\big({-Ze^2/kTr}\big)
\label{eqn:gr}
\end{equation}
where $Z = \pm 1$ for the He {\sc ii} ions and electrons.
Figure \ref{fig:grpaperIII} compares the generated $g(r)$ function for ions and electrons, at three characteristic times, with its theoretical behaviour (equation 
\ref{eqn:gr}).

The fluctuation level of the generated functions is relatively high, especially at short distances, where the number of particles contained within a thin shell centered at a distance \(r\) has a large standard deviation relative to its expected mean. Nevertheless, the three experimental curves remain consistent with one another within the fluctuation range, indicating that the particle-reinjection procedure successfully reproduces the statistical distribution of the initial particle positions.

Moreover, the behavior of the experimental curves is qualitatively consistent with the theoretical prediction, in the sense that the particle densities of both ions and electrons deviate from the uniform density at short distances in a manner consistent with their respective charges. There remains, however, a systematic bias between the theoretical and experimental curves, most notably for the electrons, although the significance of this trend is difficult to assess because of the high noise level in the experimental data.

\begin{figure}[t]
\centering
\includegraphics[width=1\linewidth]
{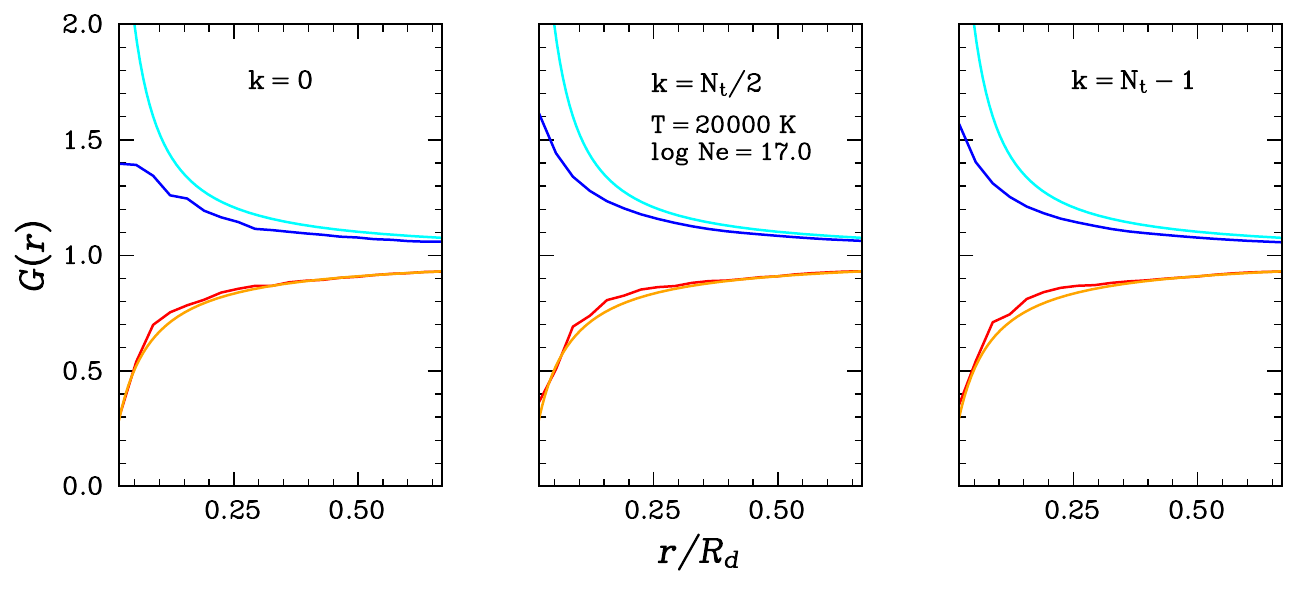}
\caption[Statistical distribution of $G(r)$ for ions and electrons]{ Statistical distribution of $G(r)$ for ions and electrons 
for a temperature of 20,000 K and
  an electronic density of $10^{17}$ cm$^{-3}$ at three time-step
  indices $k=t_k/\Delta_t$ of 0, $N_t/2$, and $N_t-1$. The distance $r$ is normalized by the Debye radius $R_d$. 
  The theoretical ion and electron functions are shown in orange and cyan, respectively, whereas the corresponding generated functions are depicted in red and blue.}
\label{fig:GrpaperIII}
\end{figure}

A way to reduce the noise level—at the cost of losing information at specific distances \(r\)—is to evaluate the statistical distribution of particles contained within a volume of radius \(r\). We define a function \(G(r)\) as the number of particles inside this radius, normalized by the number expected for a uniform distribution:
\begin{equation}
G(r) \equiv {\int_0^r dx\ g(x)x^2 \over \int_0^r dx\ x^2}= 
{1\over 3 r^3} \int_0^r dx\ g(x)x^2.
\end{equation}
The \(G(r)\) functions obtained for ions and electrons, shown in Figure~\ref{fig:GrpaperIII}, exhibit a high degree of smoothness, making it possible to clearly identify the differences between the experimental and theoretical distributions.

A slight discrepancy therefore remains, particularly for electrons. This effect may result from the minimum cut-off imposed on the point of closest approach $\sigma$ of the trajectories. Excluding trajectories with small $\sigma$ could lead to a lower-than-expected number of electrons located at short distances \(r\) from the emitter. This outcome is reminiscent of a similar result reported by \citet{Stambulchik2022} for electrons by excluding "bound" electrons from the statistics.

The preceding tests indicate that the statistical distributions agree with the expectations of statistical mechanics from the very beginning of the simulation, aside from a small, still not fully understood deviation in the spatial distribution of the electrons. As these distributions remain stable over time, no thermalization period is required to ensure that the dynamical behavior of the perturbers is statistically stationary.

\subsection{Linear Stark Effect}\label{subsec:linear}

An additional validity test of the simulation environment is the numerical verification that the linear Stark broadening characteristic of hydrogenic ions under the no–quenching approximation is correctly recovered. Owing to the degeneracy of energy levels sharing the same principal quantum number $n$, together with the particular analytical form of the dipole–transition matrix elements between states of identical $n$, the Stark effect is known to be linear for hydrogenic ions when mixing between states of different principal quantum numbers is neglected.

\begin{figure}[t]
\centering
\includegraphics[width=1\linewidth]
{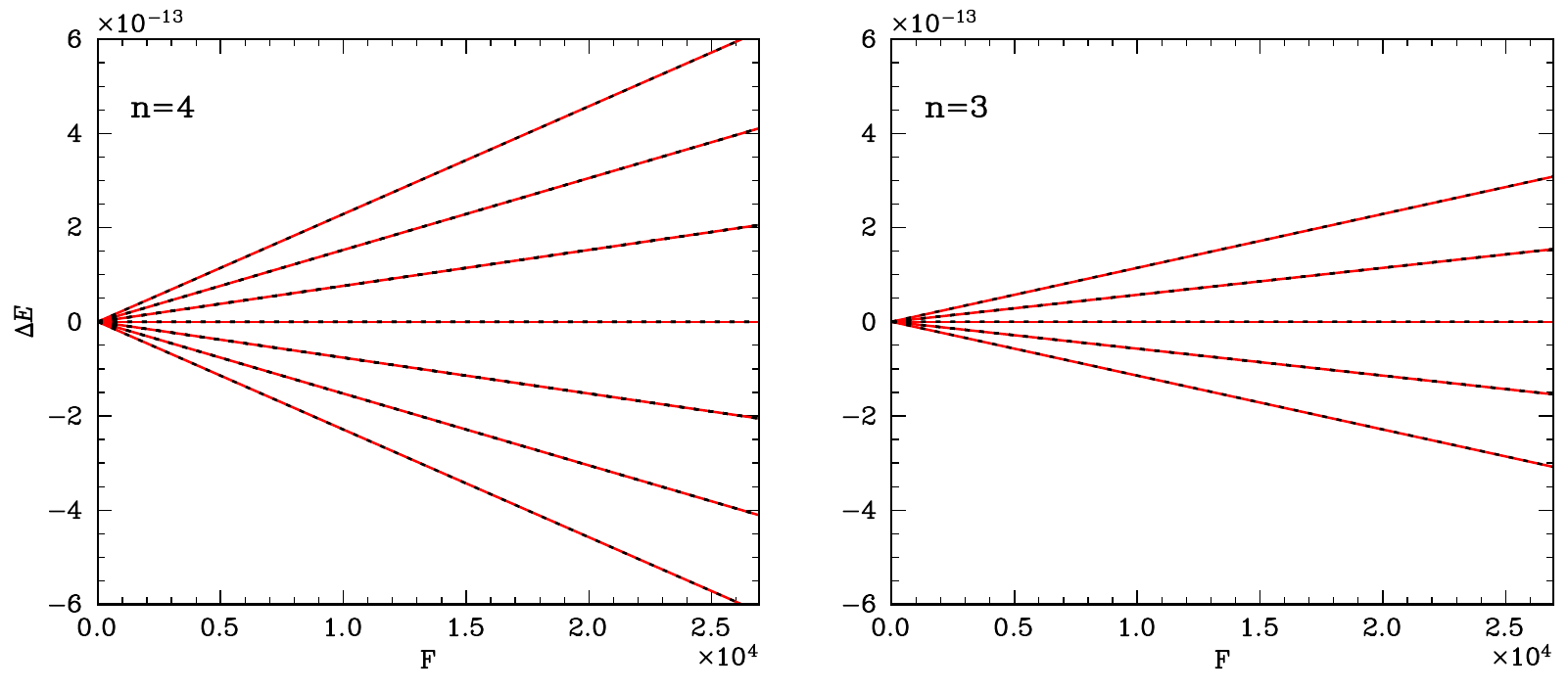}
\caption[Energy shift of the Stark components involved in the He\,\textsc{ii} $\lambda4686$ transition]
{Energy shifts of the Stark components involved in the He\,\textsc{ii} $\lambda4686$ transition. The red solid lines indicate the theoretical values for the upper ($n=4$) and lower ($n=3$) levels, as indicated in each panel, while the black dashed lines represent the values obtained from the simulation. Note that the energy shift vanishes for one Stark component of both the upper and lower levels. Quantities are expressed in CGS units.
\label{fig:figeigen}}
\end{figure}

Assuming a static uniform electric field of fixed magnitude $F$ directed along the $z$ axis, the eigenstates of the corresponding Hamiltonian for states with principal quantum number $n$ yield Stark components whose energy shifts can be described using the parabolic quantum numbers $(n_1, n_2, m)$, with the constraint
\begin{equation}
n = n_1 + n_2 + |m| + 1.
\end{equation}
Defining $q = n_{1} - n_{2}$, the energy shift of a state 
$\lvert n, n_{1}, n_{2}, m \rangle$ varies linearly with the electric field $F$ and is given by
\begin{equation}
    \Delta E_{n,q} = \frac{3}{2Z_c}\, n q\, e a_{0} F,
\end{equation}
where $Z_c$ is the core charge of the emitter. The allowed values of $q$ are the integers in the range 
$\pm (n-1)$.

Therefore, for the He\textsc{ii} $\lambda 4686$ line ($Z_c=2$), the Stark shifts of the upper and lower levels are
\[
\Delta E_{4,q_4} = 3q_4\, e a_0 F,
\qquad
q_4 = 0,\pm1,\pm2,\pm3,
\]
\[
\Delta E_{3,q_3} = \frac{9}{4}q_3\, e a_0 F,
\qquad
q_3 = 0,\pm1,\pm2.
\]

The behavior of the perturbed energies predicted by the simulation is shown in Figure~\ref{fig:figeigen} for the upper and lower levels of the He\textsc{ii} $\lambda4686$ transition as a function of the electric field. All Stark levels display the characteristic linear behavior, and there are 7 and 5 distinct Stark components for the upper and lower states, respectively, as expected.

\subsection{Hydrogen Line Profiles}\label{subsec:profH}

A final test used to validate the numerical simulation as a whole consists in computing a hydrogen line profile and comparing it with profiles from the literature.
To this end, we generated 20{,}000 electric-field sequences for a temperature of $20{,}000\,\mathrm{K}$ and an electron density of $N_e =10^{16}$ cm$^{-3}$, for which the perturbers—electrons and protons—follow the straight-line trajectories described in Section \ref{sec:line}. 
A second series of sequences was also generated in which the ions were kept static.

\begin{figure}[t]
\centering
\includegraphics[width=1\linewidth]
{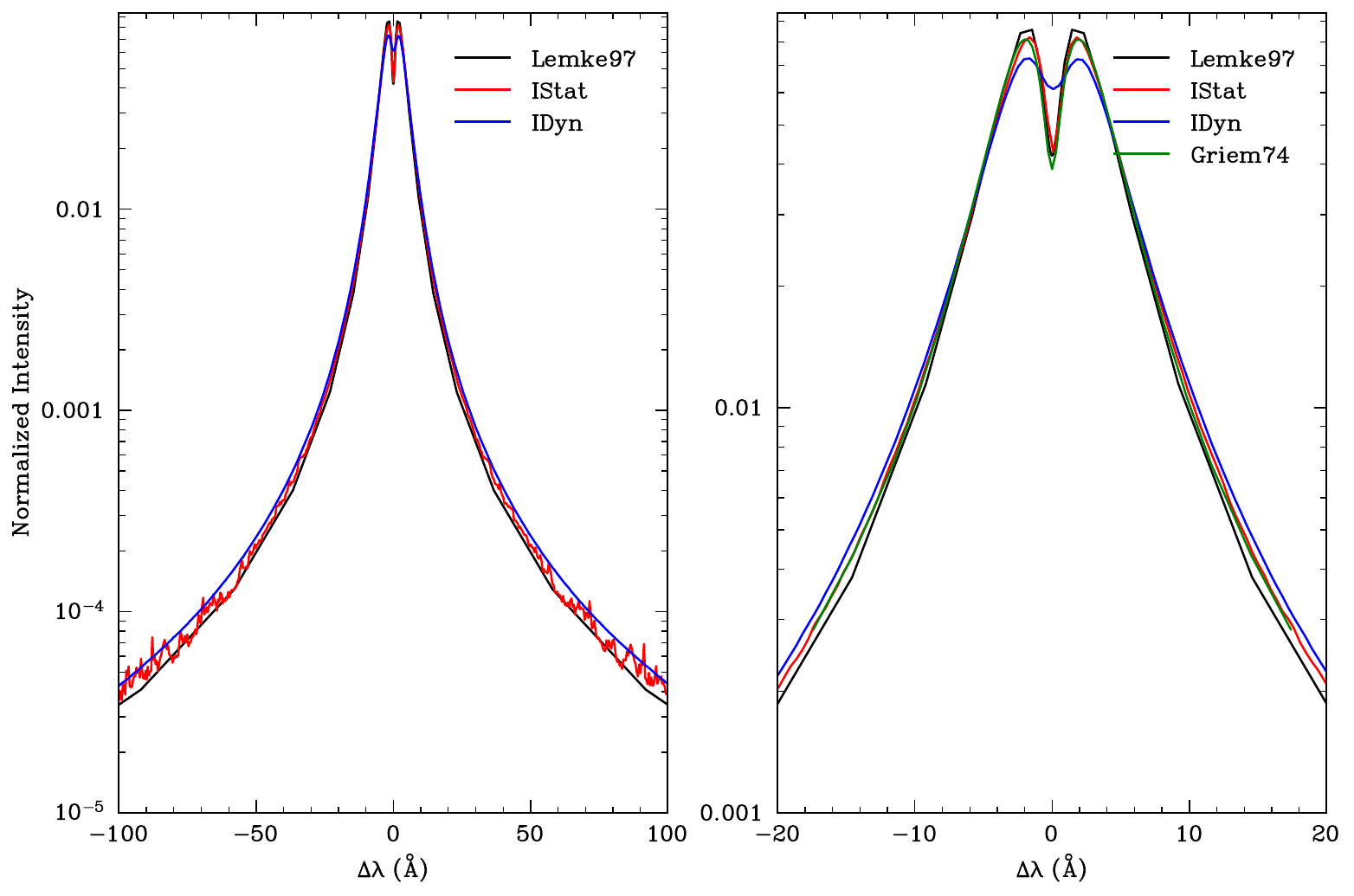}
\caption[Line profiles of the H$\beta$ $\lambda$4861  transition at an electron density of 10$^{16}$ cm$^{-3}$.]{Comparison of H$\beta$ $\lambda$4861 line profiles for a temperature of 20,000 K and an electron density of 10$^{16}$ cm$^{-3}$, from \citet[][Griem74]{Griem74}, \citet[][Lemke97]{Lemke97}, and this work, with quasi-static ions (IStat) and including ion dynamics (IDyn). The right panel is the same as the left panel but with a higher resolution in wavelength.
}
\label{fig:fig4861_1}
\end{figure}

Figure \ref{fig:fig4861_1} compares our computed H$\beta$ $\lambda$4861 line profiles —obtained using both static and dynamic ion treatments— 
and the profiles predicted by the semi‑analytical models of \citet{Griem74} and \citet{Lemke97}, in which the ionic perturbers are assumed to behave quasi‑statically.
Differences are observed in the line core and far out in the wings
between our profile including ion dynamics and those reported in the literature. 
Ion dynamics contribute directly only to the core of the line profile, but the resulting change in shape affects the total area of the profile prior to normalization; the normalization procedure, which enforces unit area, can therefore produce a global modification of the profile 
intensity, even in the far wings.

Imposing static ions in the computer simulation significantly improves the agreement. 
The residual noise in the far wings of our 
static-ion profile might arise from an insufficient number of electric-field sequences, 
which does not adequately sample the range of static ion configurations.

\subsection{He II Line Profiles}\label{subsec:prof}

The current work focuses on two first illustrative calculations. Although a full grid of 
He\,\textsc{ii}
\(\lambda\,4686\) line profiles spanning a broad range of thermodynamic conditions is planned, only two profiles have been computed so far, for a temperature of \(20{,}000\,\mathrm{K}\) and electron densities of $N_e=10^{16}$ and 10$^{17}$ cm$^{-3}$. The perturber population consists of electrons and He\,\textsc{ii} ions in equal proportions.

The principal parameters of the model are the time step \(\Delta_t\), the total number of time steps \(N_t\), the number of independent time sequences of electric fields used to compute the final profile, the minimal allowed distance from the emitter, and the radius \(R\) of the simulation volume—conveniently expressed as a multiple \(K_D\) of the Debye radius  
\[
R = K_D \sqrt{\frac{kT}{4\pi\,(N_e + N_{\mathrm{He\,II}})\,e^2}}\, .
\]

For the specific profiles discussed in this section, \(20{,}000\) independent temporal sequences of electric fields were generated, each consisting of \(N_t = 50{,}000\) values sampled at regular times \(t_k = k\,\Delta_t\). The time step was set to a fraction \(\epsilon = 0.02\) of the characteristic timescale associated with the temporal variation of the microfield,
\[
\label{eq:deltt}
\Delta_t = \epsilon\,\frac{r_0}{v_T},
\]
where \(r_0\) denotes the typical interparticle distance in a plasma of electron density \(N_e\), and \(v_T\) is the electron thermal velocity computed using the reduced mass \(\mu_e\).
The resulting duration, $N_t\,\Delta_t$, ensures that the autocorrelation function of the emitter's dipole moment has 
vanished, fulfilling one of the parameter‑validity criteria.
The radius of the simulation volume was chosen such that it contains 600 and 250 perturbers of each species, corresponding to \(K_D = 3.0\) and \(K_D \approx 3.35\), for $N_e=10^{16}$ and 10$^{17}$ cm$^{-3}$, respectively.
Trajectories for which the minimum approach distance \(\sigma\) (equation \ref{eqn:sigma}) falls below the characteristic radius of the upper level --- \(n^{2} a_{0}/Z_c\), with \(n = 4\), \(Z_c = 2\) the charge of the nucleus, and \(a_{0}\) the Bohr radius, were excluded from the simulation and replaced with admissible trajectories.

Available profile tables for the He\,\textsc{ii} $\lambda4686$ line are scarce and originate from calculations performed approximately three decades ago. The profiles published by different authors were all obtained using semi-analytical theories assuming quasi-static ions. However, as in the present work, the electrons were assumed to follow hyperbolic trajectories.

\begin{figure}[t]
\centering
\includegraphics[width=1\linewidth]
{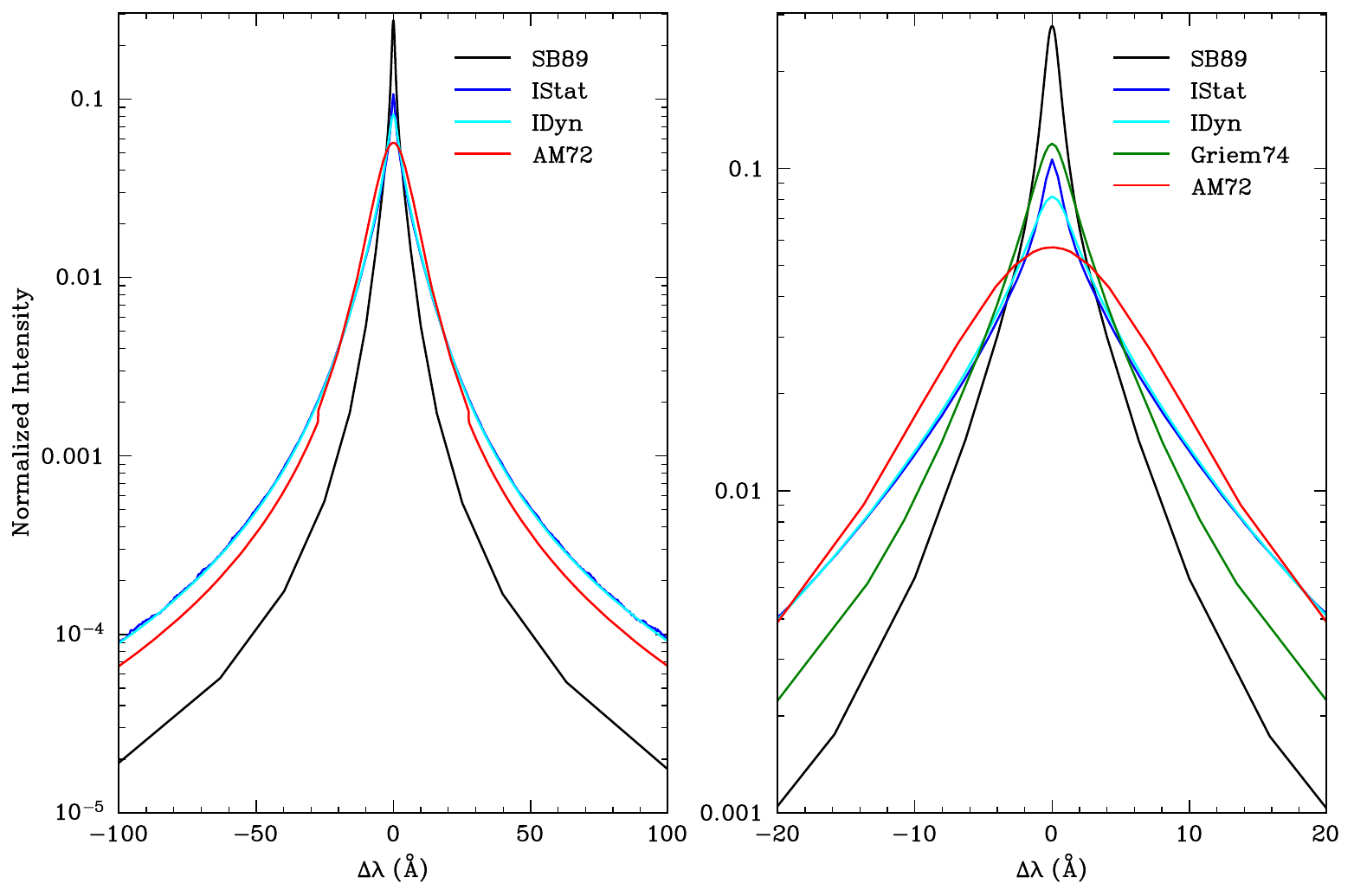}
\caption{Comparison of He {\sc ii} $\lambda$4686 line profiles for a temperature of 20,000 K and an electron density of 10$^{17}$ cm$^{-3}$, from \citet[][AM72]{AM72}, 
\citet[][Griem74, right panel only]{Griem74}, 
\citet[][SB89]{Schoening89}, and this work, with quasi-static ions (IStat) and including ion dynamics (IDyn). The right panel is the same as the left panel but with a higher resolution in wavelength.
}
\label{fig:fig4686_3}
\end{figure}

Figure \ref{fig:fig4686_3} compares the profiles computed by these
authors with the one generated in the present work for a temperature
of $20{,}000\,$K and an electron density of $N_{\mathrm{e}} =
10^{17}\,\mathrm{cm^{-3}}$. The profile from \citet{AM72} results from
our own implementation of the equations provided by the authors, which
represent an approximate formulation of the theory developed by Griem
in the 1960s.  This profile exhibits a discontinuity arising from the
transition between the electron–impact regime in the line core and the
effects of the Lewis cut-off \citep{Lewis1961} in the wings.  The
profile from \citet{Griem74} is based on the theory developed by
\citet{Kepple72}. \citet{Schoening89} rely on the Unified Theory,
which models the transition from the impact to the quasi-static regime
for electrons while keeping the ions quasi-static. Effects of ion
dynamics are therefore neglected in all of these profiles.

\begin{figure}[t]
\centering
\includegraphics[width=1\linewidth]
{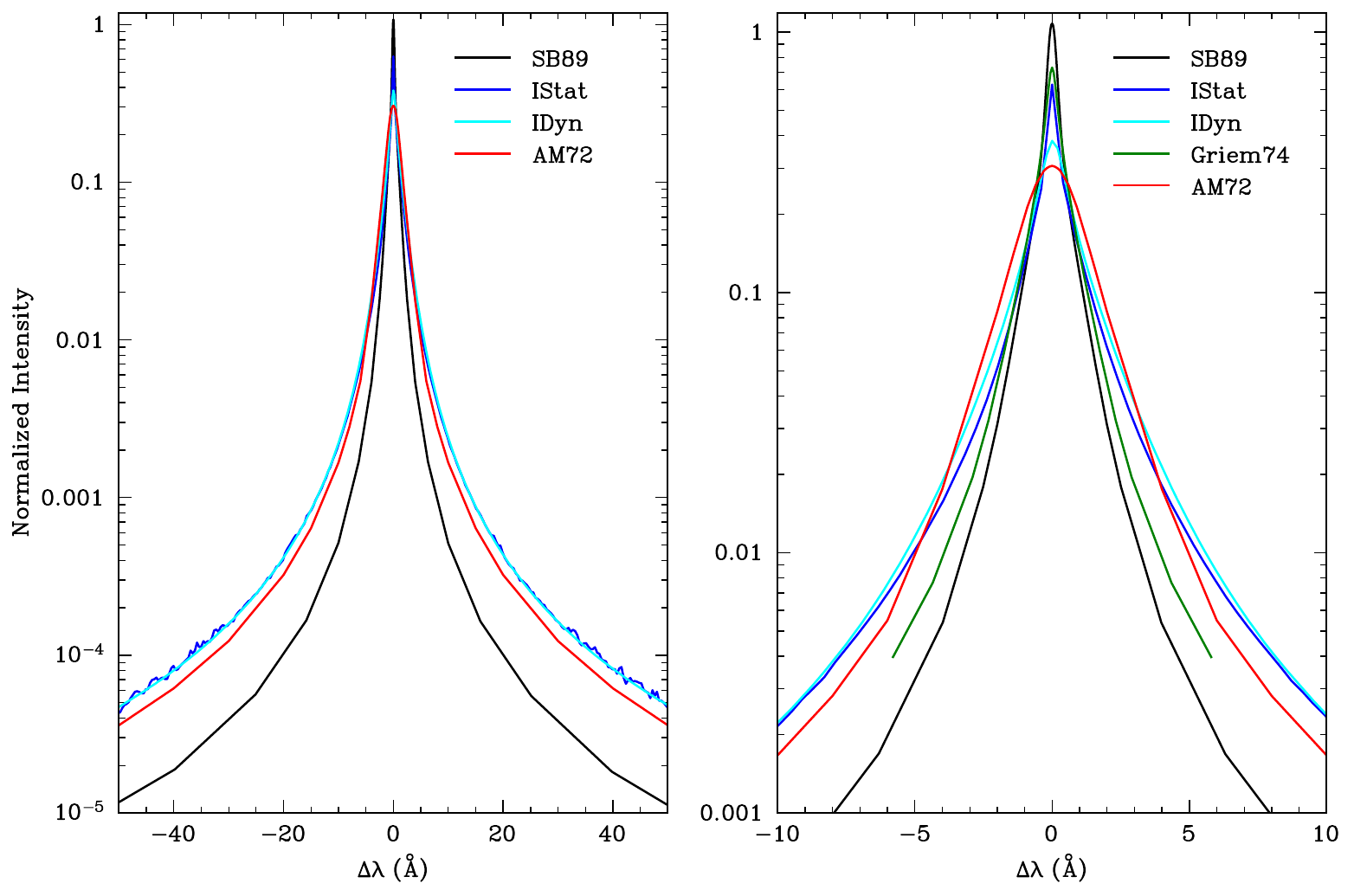}
\caption{Same as Figure \ref {fig:fig4686_3} but for a temperature of 20,000 K and an electron density of 10$^{16}$ cm$^{-3}$.
}
\label{fig:fig4686_4}
\end{figure}

Significant differences are observed both among the published profiles
and between each of them and our own. The discrepancies with our
profile cannot be attributed solely to ion dynamics, which are
included only in our simulation approach. Indeed, we generated a
second profile in which the ions were held fixed (with only the
electrons in motion), also shown in the figure. The contribution of
ion dynamics is negligible at this density, consistent with our
results for neutral helium \citep{Tremblay2026}.  Similar discrepancies
are observed at a density of $N_e = 10^{16}$ cm$^{-3}$, as illustrated
in Figure \ref{fig:fig4686_4}.

\section{Discussion}

The Stark-broadened profiles we have computed for the He {\sc ii} $\lambda$4686 transition differ in several important respects from the profiles currently used in all published model-atmosphere analyses of DO white dwarfs. These differences arise directly from the improved physical treatment implemented in our simulations—most notably the explicit inclusion of a large set of electric-field time sequences, which allows accurate integration of the ion’s time‑dependent Hamiltonian.

At this stage, our simulation environment successfully passes all validation tests, 
including the statistical distribution of particle coordinates, the distribution of 
the electric field, the expected linear behavior of the Stark component energies, 
and the calculation of an H$\beta$ $\lambda 4861$ line profile at a specific 
temperature and density that agrees with the literature.
Our attempts to explain the remaining discrepancies have been unsuccessful. 
This contrasts with our results for the neutral-helium profiles
\citep{Tremblay2026}, which are nearly identical whether obtained from the semi-analytical theory or from the numerical simulation, except for the contribution of ion dynamics at lower densities.

The discrepancies we find between our new profiles and the earlier ones, particularly those of \citet{Schoening89}, are not merely of academic interest. As discussed in the Introduction, spectroscopic analyses of hot DO white dwarfs have long exhibited puzzling and physically implausible trends: inferred surface gravities that increase sharply at high effective temperatures, mass distributions that differ markedly from those of cooler DB stars, and systematic offsets relative to the DA sequence. These anomalies have resisted explanation despite extensive efforts to improve atmospheric models, cooling sequences, and fitting techniques. Given that He {\sc ii} $\lambda$4686 is the dominant diagnostic feature in DO spectra, it is reasonable to suspect that inaccuracies in the underlying Stark broadening theory may be a significant—perhaps even primary—contributor to these long-standing issues.

Our exploratory calculations therefore raise the intriguing possibility that the spectroscopic mass discrepancies observed in DO white dwarfs may originate, at least in part, from limitations in the Stark profiles that have been used for decades. However, it is essential to emphasize that this conclusion cannot yet be drawn with confidence. The present work focuses exclusively on a single transition, and only for a limited set of plasma conditions. To assess the true impact of our improved treatment, a complete grid of He {\sc ii} line profiles must be computed and incorporated into full non-LTE model atmospheres. Only then will it be possible to perform a systematic reanalysis of DO spectra and determine whether the new profiles lead to more physically consistent atmospheric parameters.

In this sense, the results presented here should be viewed as a first step—an encouraging indication that the standard Stark broadening theory for ionized helium may require revision, but not yet a definitive solution to the DO mass problem. The next phase of this project will involve extending our simulations to all relevant He {\sc ii} transitions and generating a comprehensive set of profiles suitable for model-atmosphere calculations. Once this grid is available, we will be able to test directly whether our improved Stark profiles yield more coherent mass distributions and eliminate the anomalous trends observed in previous spectroscopic studies. A similar study performed with neutral-helium profiles and a large sample of DB white dwarfs has indeed revealed avenues for improving our profile calculations with the semi-analytical theory  \citep{Tremblay2026b}.

If these forthcoming analyses confirm that the new profiles resolve the inconsistencies highlighted in the Introduction, this would represent a significant advance in our understanding of DO white dwarfs and, more broadly, in the modeling of dense helium plasmas. Until then, the results presented here should be interpreted as a strong motivation for further work rather than a final verdict.

\appendix

\section[\appendixname~\thesection]{Jacobian of the Hyperbolic Impact Coordinate Transformation: Mathematical Proof \label{sec:determ2}}

The purpose of this section is to derive the jacobian of $J$ (equation \ref{eqn:Jmatrix}).
For clarity, \(v\) is taken to be \(v_\infty\) in the derivation.
Moreover, unless explicitly stated otherwise, the matrices presented in this section are square of dimension \(6 \times 6\), and their elements are sometimes expressed as transposed (horizontal) 3D-vectors in order to lighten the notation.

Based on the results of the Section \ref{subsec:dynamic}, the position and velocity of a particle on a hyperbolic trajectory at time \( t \) can be written as  
\begin{eqnarray}
{\bf r}(t) &=& X_1{\bf w_1} + X_2{\bf w_2}\nonumber \\
{\bf v}(t) &=& V_1{\bf w_1} + V_2{\bf w_2},
\end{eqnarray}  

\noindent
where the orthonormal basis \( {\bf w_{1,2,3}} \) is defined in equations (\ref{eqn:w3}) and (\ref{eqn:w12}), and the coefficients \( (X_{1,2},V_{1,2}) \) are functions of the velocity at infinity \( v \), the impact parameter \( b \), and the time of closest approach \( \tau_b \):  
\begin{eqnarray}
X_1 &=& a(\cosh(u)+\lambda e)\nonumber\\
X_2 &=&  b\sinh(u) \nonumber\\
V_1 &=& \sqrt{\kappa\over a^3}{1\over e\cosh(u)+\lambda} a\sinh(u)\nonumber \\
V_2 &=&  \sqrt{\kappa\over a^3}{1\over e\cosh(u)+\lambda}b\cosh(u),
\label{eqn:x12v12}
\end{eqnarray}  

\noindent
where \( a \), \( e \), and \( u \) are functions of the coordinates:  
\begin{eqnarray} 
a(v) &=& {\kappa\over v^2}\nonumber \\ 
e(a,b) &=& {\sqrt{a^2(v) + b^2}\over a(v)}\nonumber \\ 
{\sqrt{a^2(v) + b^2}\over a(v)}\sinh(u) + \lambda u &=& \sqrt{\kappa\over a^3(v)}(t-\tau_b). 
\end{eqnarray}  

As discussed for straight-line trajectories (equation \ref{eqn:pxv3}), the probability density of the particle's position in the new coordinates is given by the product of a Boltzmann factor and the jacobian of the coordinate transformation. For hyperbolic trajectories, the jacobian matrix \( {\bf J} \) is expressed as  
\begin{equation}
{\bf J} \equiv \left (\begin{matrix}
{\partial {\bf r^\text{T}} \over \partial v} & {\partial {\bf v^\text{T}} \over \partial v} \\
{\partial {\bf r^\text{T}} \over \partial b} &{\partial {\bf v^\text{T}} \over \partial b}  \\
{\partial {\bf r^\text{T}} \over \partial \tau_b} &{\partial {\bf v^\text{T}} \over \partial \tau_b}  \\
{\partial {\bf r^\text{T}} \over \partial \theta} &{\partial {\bf v^\text{T}} \over \partial \theta}  \\
{\partial {\bf r^\text{T}} \over \partial \phi} &{\partial {\bf v^\text{T}} \over \partial \phi}  \\
{\partial {\bf r^\text{T}} \over \partial \beta} &{\partial {\bf v^\text{T}} \over \partial \beta}  \end{matrix}
\right ).
\end{equation}  

\noindent
Following the approach used for straight-line trajectories, all elements of \( {\bf J} \) can be expressed as a linear expansion in the orthonormal basis \( {\bf w_{1,2,3}^\text{T}} \) to simplify calculations. The first three rows explicitly follow this form (with zero weighting for \( {\bf w_3^\text{T}} \)), whereas rows 4 to 6 do not:  

\begin{equation}
{\bf J} = \left (\begin{matrix}
{\partial X_1 \over \partial v}{\bf w_1^\text{T}}+{\partial X_2 \over \partial v}{\bf w_2^\text{T}}
 & 
{\partial V_1 \over \partial v}{\bf w_1^\text{T}}+{\partial V_2 \over \partial v}{\bf w_2^\text{T}}
\\
{\partial X_1 \over \partial b}{\bf w_1^\text{T}}+{\partial X_2 \over \partial b}{\bf w_2^\text{T}}
 & 
{\partial V_1 \over \partial b}{\bf w_1^\text{T}}+{\partial V_2 \over \partial b}{\bf w_2^\text{T}}
\\
{\partial X_1 \over \partial \tau_b}{\bf w_1^\text{T}}+{\partial X_2 \over \partial \tau_b}{\bf w_2^\text{T}}
 & 
{\partial V_1 \over \partial \tau_b}{\bf w_1^\text{T}}+{\partial V_2 \over \partial \tau_b}{\bf w_2^\text{T}}
\\
X_1{\partial {\bf w_1^\text{T}} \over \partial \theta}+X_2{\partial {\bf w_2^\text{T} }\over \partial \theta}
 & 
V_1{\partial {\bf w_1^\text{T}} \over \partial \theta}+V_2{\partial {\bf w_2^\text{T} }\over \partial \theta}
\\
X_1{\partial {\bf w_1^\text{T}} \over \partial \phi}+X_2{\partial {\bf w_2^\text{T} }\over \partial \phi}
 & 
V_1{\partial {\bf w_1^\text{T}} \over \partial \phi}+V_2{\partial {\bf w_2^\text{T} }\over \partial \phi}
\\
X_1{\partial {\bf w_1^\text{T}} \over \partial \beta}+X_2{\partial {\bf w_2^\text{T} }\over \partial \beta}
 & 
V_1{\partial {\bf w_1^\text{T}} \over \partial \beta}+V_2{\partial {\bf w_2^\text{T} }\over \partial \beta}
\end{matrix}
\right ).
\end{equation}  

\noindent
Because any vector can be expanded in this basis, a set of coefficients \((P, Q, R, p, q, r)_{1,2,3}\) can be introduced as follows:
\begin{eqnarray}
X_1{\partial {\bf w_1^\text{T}} \over \partial \theta} + X_2{\partial {\bf w_2^\text{T}} \over \partial \theta}
&\equiv& P_1{\bf w_1^\text{T}} + P_2{\bf w_2^\text{T}} + P_3{\bf w_3^\text{T}}, \nonumber \\
X_1{\partial {\bf w_1^\text{T}} \over \partial \phi} + X_2{\partial {\bf w_2^\text{T}} \over \partial \phi}
&\equiv& Q_1{\bf w_1^\text{T}} + Q_2{\bf w_2^\text{T}} + Q_3{\bf w_3^\text{T}}, \nonumber \\
X_1{\partial {\bf w_1^\text{T}} \over \partial \tau_b} + X_2{\partial {\bf w_2^\text{T}} \over \partial \tau_b}
&\equiv& R_1{\bf w_1^\text{T}} + R_2{\bf w_2^\text{T}} + R_3{\bf w_3^\text{T}}, \nonumber \\
V_1{\partial {\bf w_1^\text{T}} \over \partial \theta} + V_2{\partial {\bf w_2^\text{T}} \over \partial \theta}
&\equiv& p_1{\bf w_1^\text{T}} + p_2{\bf w_2^\text{T}} + p_3{\bf w_3^\text{T}}, \nonumber \\
V_1{\partial {\bf w_1^\text{T}} \over \partial \phi} + V_2{\partial {\bf w_2^\text{T}} \over \partial \phi}
&\equiv& q_1{\bf w_1^\text{T}} + q_2{\bf w_2^\text{T}} + q_3{\bf w_3^\text{T}}, \nonumber \\
V_1{\partial {\bf w_1^\text{T}} \over \partial \beta} + V_2{\partial {\bf w_2^\text{T}} \over \partial \beta}
&\equiv& r_1{\bf w_1^\text{T}} + r_2{\bf w_2^\text{T}} + r_3{\bf w_3^\text{T}}.
\end{eqnarray}

\noindent
The jacobian matrix can now be written as the product of two simpler matrices:

\begin{equation}
{\bf J} = 
\left (\begin{matrix}
{\partial X_1 \over \partial v} & {\partial X_2 \over \partial v} & 0
& {\partial V_1 \over \partial v} & {\partial V_2 \over \partial v} & 0 \\
{\partial X_1 \over \partial b} & {\partial X_2 \over \partial b} & 0 
& {\partial V_1 \over \partial b} & {\partial V_2 \over \partial b} & 0 \\
{\partial X_1 \over \partial t_0} & {\partial X_2 \over \partial t_0} & 0 
& {\partial V_1 \over \partial t_0} & {\partial V_2 \over \partial t_0} & 0 \\
P_1 & P_2 & P_3 & p_1 & p_2 & p_3 \\
Q_1 & Q_2 & Q_3 & q_1 & q_2 & q_3 \\
R_1 & R_2 & R_3 & r_1 & r_2 & r_3 
\end{matrix} \right )
\left (\begin{matrix}
{\bf w_1^\text{T}} & {\bf 0^\text{T}} \\
{\bf w_2^\text{T}} & {\bf 0^\text{T}} \\
{\bf w_3^\text{T}} & {\bf 0^\text{T}} \\
{\bf 0^\text{T}} & {\bf w_1^\text{T}} \\
{\bf 0^\text{T}} & {\bf w_2^\text{T}} \\
{\bf 0^\text{T}} & {\bf w_3^\text{T}} 
\end{matrix} \right ).
\end{equation}

\noindent
Given that the determinant of the second matrix is unity, 
the determinant of \(\mathbf{J}\) simplifies to that of the first matrix:

\begin{equation}
J= \left |\begin{matrix}
{\partial X_1 \over \partial v} & {\partial X_2 \over \partial v} & 0
& {\partial V_1 \over \partial v} & {\partial V_2 \over \partial v} & 0 \\
{\partial X_1 \over \partial b} & {\partial X_2 \over \partial b} & 0 
& {\partial V_1 \over \partial b} & {\partial V_2 \over \partial b} & 0 \\
{\partial X_1 \over \partial \tau_b} & {\partial X_2 \over \partial \tau_b} & 0 
& {\partial V_1 \over \partial \tau_b} & {\partial V_2 \over \partial \tau_b} & 0 \\
P_1 & P_2 & P_3 & p_1 & p_2 & p_3 \\
Q_1 & Q_2 & Q_3 & q_1 & q_2 & q_3 \\
R_1 & R_2 & R_3 & r_1 & r_2 & r_3 
\end{matrix} \right |,
\end{equation}

\noindent
which is considerably simpler than the original expression. A detailed calculation further shows that
\begin{eqnarray}
R_1 &=& -X_2, \nonumber \\
R_2 &=& X_1, \nonumber \\
R_3 &=& 0, \nonumber \\
(P_3q_3 - Q_3p_3) &=& \lambda \sqrt{\kappa\over a^3} ab \sin\theta.
\end{eqnarray}

\noindent
This leads to a further simplification of the jacobian:

\begin{equation} 
J = \lambda \sqrt{\kappa\over a^3} ab \sin\theta
\left |\begin{matrix} 
{\partial X_1 \over \partial v} & {\partial X_2 \over \partial v} & {\partial V_1 \over \partial v} & {\partial V_2 \over \partial v} \\ 
{\partial X_1 \over \partial b} & {\partial X_2 \over \partial b} & {\partial V_1 \over \partial b} & {\partial V_2 \over \partial b} \\ 
{\partial X_1 \over \partial \tau_b} & {\partial X_2 \over \partial \tau_b} & {\partial V_1 \over \partial \tau_b} & {\partial V_2 \over \partial \tau_b} \\ 
-X_2 & X_1 & -V_2 & V_1 
\end{matrix} \right | .
\end{equation}

\noindent
The \(4 \times 4\) matrix above is independent of the angles \(\phi\) and \(\beta\), implying that the jacobian is proportional to \(\sin\theta\) and does not depend on these angles.

The calculation remains tedious because the quantities \(X_{1,2}\) and \(V_{1,2}\) are functions of \(a\), \(e\), and \(u\), which themselves depend nonlinearly on \(v\), \(b\), and \(\tau_b\). However, by applying the chain rule, the jacobian matrix \(\mathbf{J}\) can be expressed as the product of two matrices, each with a determinant that is easier to compute.

Let \( W \) be any of the four quantities \( X_1, X_2, V_1, V_2 \). The function \( W(v,b,\tau_b) \) can be rewritten in terms of \( a \) (which is a function of \( v \)), the eccentricity \( e \) (which depends on \( a(v) \) and \( b \)), and \( u \) (which is a function of \( a(v) \), \( b \), and \( \tau_b \); see equation \ref{eqn:x12v12}). Applying the chain rule, the partial derivatives of any \( W \) (i.e., \( X_{1,2} \) and \( V_{1,2} \)) with respect to \( v \), \( b \), and \( \tau_b \) are given by:

\begin{eqnarray}
\left(\frac{\partial W}{\partial v}\right)_{b,\tau_b} 
&=&
\left(\frac{\partial W}{\partial a}\right)_{e,u} \frac{da}{dv}
+
\left(\frac{\partial W}{\partial e}\right)_{a,u}
\left(\frac{\partial e}{\partial a}\right)_{b} \frac{da}{dv}
+
\left(\frac{\partial W}{\partial u}\right)_{a,e}
\left(\frac{\partial u}{\partial a}\right)_{b,\tau_b} \frac{da}{dv}
\nonumber \\
\left(\frac{\partial W}{\partial b}\right)_{v,\tau_b} 
&=&
\left(\frac{\partial W}{\partial e}\right)_{a,u}
\left(\frac{\partial e}{\partial b}\right)_{a}
+
\left(\frac{\partial W}{\partial u}\right)_{a,e}
\left(\frac{\partial u}{\partial b}\right)_{a,\tau_b}
\nonumber \\
\left(\frac{\partial W}{\partial \tau_b}\right)_{v,b} 
&=&
\left(\frac{\partial W}{\partial u}\right)_{a,e}
\left(\frac{\partial u}{\partial \tau_b}\right)_{a,b}.
\end{eqnarray}

\noindent
The notation, commonly used in thermodynamics, highlights the variables held constant when computing partial derivatives. This formulation allows the \( 4 \times 4 \) matrix to be rewritten as the product of two matrices. The jacobian determinant now takes the form:

\begin{eqnarray} 
J &=& \lambda \sqrt{\frac{\kappa}{a^3}} ab \sin\theta \left|
\begin{matrix} 
\frac{da}{dv} & \frac{\partial e}{\partial a} \frac{da}{dv} & \frac{\partial u}{\partial a} \frac{da}{dv} & 0\\  
0 & \frac{\partial e}{\partial b} & \frac{\partial u}{\partial b} & 0 \\  
0 & 0 & \frac{\partial u}{\partial \tau_b} & 0 \\  
0 & 0 & 0 & 1 
\end{matrix} 
\right|_{vb\tau_b} 
\left|
\begin{matrix} 
\frac{\partial X_1}{\partial a} & \frac{\partial X_2}{\partial a} & \frac{\partial V_1}{\partial a} & \frac{\partial V_2}{\partial a} \\ 
\frac{\partial X_1}{\partial e} & \frac{\partial X_2}{\partial e} & \frac{\partial V_1}{\partial e} & \frac{\partial V_2}{\partial e} \\ 
\frac{\partial X_1}{\partial u} & \frac{\partial X_2}{\partial u} & \frac{\partial V_1}{\partial u} & \frac{\partial V_2}{\partial u} \\  
- X_2 & X_1 & -V_2 & V_1 
\end{matrix}
\right|_{aeu} \nonumber \\ 
&=& 
2\lambda \frac{1}{e\cosh u + \lambda} \sqrt{\frac{\kappa}{a^3}} \frac{b^2}{ae} \sin\theta
\left|
\begin{matrix} 
\frac{\partial X_1}{\partial a} & \frac{\partial X_2}{\partial a} & \frac{\partial V_1}{\partial a} & \frac{\partial V_2}{\partial a} \\ 
\frac{\partial X_1}{\partial e} & \frac{\partial X_2}{\partial e} & \frac{\partial V_1}{\partial e} & \frac{\partial V_2}{\partial e} \\ 
\frac{\partial X_1}{\partial u} & \frac{\partial X_2}{\partial u} & \frac{\partial V_1}{\partial u} & \frac{\partial V_2}{\partial u} \\  
- X_2 & X_1 & -V_2 & V_1 
\end{matrix} 
\right|,
\end{eqnarray} 

\noindent
using the following partial derivatives:

\begin{eqnarray} 
 \frac{da}{dv} &=& -2\sqrt{\frac{a^3}{\kappa}} \nonumber \\ 
 \left(\frac{\partial e}{\partial b}\right)_a &=& \frac{b}{a^2 e} \nonumber \\ 
\left(\frac{\partial u}{\partial \tau_b}\right)_{ab} &=& -\sqrt{\frac{\kappa}{a^3}} \frac{1}{e\cosh u + \lambda}. 
\end{eqnarray}

\noindent
Evaluating the remaining determinant leads to the jacobian:
\begin{equation} 
J = bv^3 \sin\theta.
\end{equation}
This calculation is performed by partitioning the last matrix into four $2\times 2$ blocks and applying the Schur complement formula.

\section[\appendixname~\thesection]{Random Generation of Coordinates at Initialization}\label{app:I}

In this section, $v_{\infty}$ is denoted by $v$ for notational simplicity.

Each particle in the simulation requires six initial parameters to fully describe its trajectory:
$(b, v, \tau_b, \alpha, \theta, \phi)$. 
While the angular variables can be generated straightforwardly using a uniform random number generator, this is not the case for the parameters $(b, v, \tau_b)$ associated with hyperbolic trajectories. The difficulty is twofold: first, the complexity of the underlying equations prevents the analytical inversion required to directly extract $(b, v, \tau_b)$ from uniformly distributed random numbers; second, the dependence of $b$ on $v$ imposes a specific ordering in the generation of these parameters.

Since the distribution of $v$ does not depend on any previously generated quantity (see equation \ref{eqn:bayes2}), it is generated first and subsequently used in the determination of $b$ and $\tau_b$. To sample the velocity $v$, we tabulate the cumulative integral
\begin{equation}
I(v) \equiv K_V^{-1} \int_{v_{\rm min}}^v dv'\,
e^{-v'^2 / v_T^2} v'^2
\left[
I_1\!\left(R, \lambda \frac{\kappa}{v'^2}\right)
+ \lambda \frac{\kappa}{v'^2}
I_2\!\left(R, \lambda \frac{\kappa}{v'^2}\right)
\right],
\end{equation}
for a sufficiently dense set of values between $v_{\rm min}$ (defined in equation \ref{eqn:vmin}) and approximately $5v_T$, where $v_T$ denotes the thermal velocity. By construction, $I(v) \in [0,1]$, and each value of the integral corresponds uniquely to a velocity $v$. A cubic spline interpolation is therefore constructed to relate a uniformly distributed random number $u \in [0,1]$ to the corresponding velocity $v$.

Once $v$ has been generated, the impact parameter $b$ is obtained by solving
\begin{eqnarray}
I(b) &=& \int_0^b db'\, p(b'|v) \nonumber \\
&=& K_B^{-1} \int_0^b db'\, b'
\left(
\sqrt{(R-\lambda a)^2 - (a^2+b'^2)}
+ \lambda a \cosh^{-1}\!\left(\frac{R-\lambda a}{\sqrt{a^2+b'^2}}\right)
\right),
\label{eqn:intb}
\end{eqnarray}
where $a$ is a function of the previously generated velocity $v$.

Since $I(b)$ is available in analytical form, the solution for $b$ corresponding to a given random number $u \in [0,1]$ can be obtained using a Newton-Raphson iterative scheme:
\begin{equation}
b_{k+1} = b_k - \frac{I(b_k) - u}{I'(b_k)}.
\end{equation}

Using equation (\ref{eqn:bvH32}), we write
\begin{equation}
I'(b) = K_B^{-1} F'(b),
\end{equation}
with
\begin{equation}
F'(b) =
b \left(
\sqrt{(R-\lambda a)^2 - (a^2+b^2)}
+ \lambda a \cosh^{-1}\!\left(\frac{R-\lambda a}{\sqrt{a^2+b^2}}\right)
\right),
\end{equation}
and
\begin{equation}
I(b) = K_B^{-1}\big(F(b) - F(0)\big),
\end{equation}
where
\begin{eqnarray}
F(b) &=& -\frac{1}{3}\big(R^2 - 2R\lambda a - b^2\big)^{3/2} \nonumber \\
&& -\frac{1}{2}\lambda a (R-\lambda a)
\sqrt{(R-\lambda a)^2 - (a^2+b^2)} \nonumber \\
&& +\frac{1}{2}\lambda a
\cosh^{-1}\!\left(\frac{R-\lambda a}{\sqrt{a^2+b^2}}\right)
(a^2+b^2).
\end{eqnarray}

\noindent The normalization constant is therefore defined as
\begin{equation}
K_B = F(b_{\rm max}) - F(0).
\end{equation}

\noindent With these definitions, the Newton--Raphson iteration becomes
\begin{equation}
b_{k+1} =
b_k -
\frac{\big(F(b_k) - F(0)\big) - \big(F(b_{\rm max}) - F(0)\big)u}
{F'(b_k)}.
\end{equation}

To ensure convergence, an appropriate initial guess $b_0$ must be provided. Numerical inspection of $I(b)$ as a function of $b/b_{\rm max}$ shows that it depends only weakly on $R$, $a$, and $\lambda$. Moreover, it closely resembles the analytical expression obtained in the limiting case $a=0$:
\begin{equation}
I(b; a=0)
= 1 - R^{-3}(R^2 - b^2)^{3/2}
= 1 - \left(1 - \frac{b^2}{b_{\rm max}^2}\right)^{3/2},
\end{equation}
where $b_{\rm max} = R$ for $a=0$. This approximation provides the initial estimate
\begin{equation}
b_0 = b_{\rm max} \sqrt{1 - (1-u)^{2/3}}.
\end{equation}

Extensive numerical tests over a broad range of values of $R$, $a$, and $u \in [0,1]$, and for both signs of $\lambda$, indicate that the procedure converges within at most five iterations, achieving an accuracy better than $10^{-12}$.

Finally, once $v$ and $b$ have been determined, the time of closest approach $\tau_b$ is obtained by drawing a random number uniformly distributed in the interval $[-t_{\rm max},\, t_{\rm max}]$, with $t_{\rm max}$
given by equation (\ref{eqn:tmaxH}).

\section[\appendixname~\thesection]{Numerical Solution of the Hyperbolic Eccentric Anomaly Equation \label{sec:AppHyper}}

Once the hyperbolic trajectories of the particles have been characterized by the set of parameters $(b, v, \tau_b, \alpha, \theta, \phi)$, their temporal evolution becomes one of the main challenges of the numerical simulation.

Unlike the linear trajectory around a neutral emitter, a hyperbolic trajectory around a charged emitter does not follow a fixed axis. Moreover, the position of a particle along its trajectory evolves nonlinearly in time. Consequently, the evolution must be computed individually for each particle by relating the simulation time $t$ to the hyperbolic anomaly $u$, which describes the position along the trajectory, through the nonlinear equation (\ref{eqn:Fi}). Two numerical approaches may be employed to solve this equation: the Newton-Raphson iterative method, previously used for the determination of the impact parameter $b$, and the cubic spline interpolation method used for the velocity parameter $v$.

Each method presents advantages and limitations. However, because the temporal evolution is nonlinear and must be evaluated independently for every particle in the simulation, the Newton-Raphson method is not retained. This choice increases the demand on volatile memory (RAM), but reduces the overall computational time. It should be noted that, depending on the architecture of the computing cluster, one method may outperform the other. In this Appendix, we focus on the cubic spline interpolation approach.

Since the evolution of the system is determined through the velocity of each perturber, it is necessary to ensure that each trajectory is adequately sampled within the interval $[-t_{\rm max},\, t_{\rm max}]$, as defined in equation (\ref{eqn:tmaxH}). To this end, we use
\begin{equation}
e\,\sinh(u) + \lambda u = \sqrt{\frac{\kappa}{a^3}}(t - \tau_b),
\label{eqn:Annexe1}
\end{equation}
which relates the hyperbolic anomaly $u$ to time $t$. To determine the bounds of the anomaly, in particular $u_{\rm max}$, we assume that the perturber enters the simulation volume at its boundary, thereby ensuring complete coverage of the trajectory within the domain. Under this assumption, equation (\ref{eqn:Annexe1}) becomes
\begin{equation}
e\ \sinh(u_{\rm max}) + \lambda u_{\rm max} = \sqrt{\frac{\kappa}{a^3}}\, t_{\rm max}.
\label{eqn:Annexe2}
\end{equation}
Solving this equation numerically yields the value of $u_{\rm max}$ corresponding to the time span within the simulation volume. A sufficiently fine grid is then generated over the interval
\[
u \in [-u_{\rm max} - \varepsilon,\; u_{\rm max} + \varepsilon],
\]
where $\varepsilon$ is a small extension introduced to preserve the validity of the cubic spline interpolation $S_{u}(t)$ constructed from equation (\ref{eqn:Annexe1}).

The function $S_{u}(t)$, which is specific to each perturber, allows the simulation to advance with uniform time stepping while preserving the intrinsic nonlinear dependence of the hyperbolic anomaly $u$. Once a perturber exits the simulation volume, a new spline function $S_{u}(t)$ is generated using the updated impact parameter $b$ and velocity $v$ of the newly entering perturber.

\section[\appendixname~\thesection]{Random Generation of Incoming Particles \label{app:III}}

In this section, $v_\infty$ is denoted by $v$ in order to simplify the notation.

The generation of incoming particles follows a procedure similar to that used for the initialization of particles. However, since the probability distribution for incoming particles is proportional to the initialization distribution divided by the crossing time $t_{\rm max}$, several expressions simplify accordingly.

To determine the velocity at infinity, $v$, of each incoming particle, we tabulate the cumulative integral
\begin{eqnarray}
I(v) &\equiv& K_V^{-1} \int_{v_{\rm min}}^v dv'\,
e^{-v'^2/v_T^2}\, v'
\left(v'^2 - 2\lambda \frac{\kappa}{R}\right)
\label{eqnIv2} \\
\nonumber \\
&=&
K_V^{-1}
\frac{v_T^4}{2}\,
e^{-v'^2/v_T^2}
\left(
-\frac{v'^2}{v_T^2} - 1
+ 2\lambda \frac{\kappa}{R v_T^2}
\right)
\Bigg|_{v_{\rm min}}^{v}.
\end{eqnarray}
The tabulation is performed over a sufficiently wide range of velocities, from $v_{\rm min}$ up to a large upper bound, typically $\sim 5v_T$, where $v_T$ denotes the thermal velocity.

The normalization constant is given by
\begin{eqnarray}
K_V &=& \int_{v_{\rm min}}^\infty dv'\,
e^{-v'^2/v_T^2}\, v'
\left(v'^2 - 2\lambda \frac{\kappa}{R}\right)
\nonumber \\
&=&
\frac{v_T^4}{2}\,
e^{-v'^2/v_T^2}
\left(
-\frac{v'^2}{v_T^2} - 1
+ 2\lambda \frac{\kappa}{R v_T^2}
\right)
\Bigg|_{v_{\rm min}}^{\infty}.
\end{eqnarray}

A velocity $v$ is then obtained by drawing a uniform random number $u \in [0,1]$ and solving
\begin{equation}
I(v) = u,
\end{equation}
which is achieved by interpolating within the precomputed grid of $I(v)$ values, using the same spline-based procedure adopted for particle initialization.

Once the velocity has been determined according to the incoming-particle distribution, the corresponding impact parameter $b$ is generated from the conditional probability density $p_{\rm in}(b|v)$. The associated cumulative distribution function is
\begin{equation}
I(b) = \int_0^b db'\, p_{\rm in}(b'|v)
= \left(\frac{b}{b_{\rm max}}\right)^2,
\label{eqn:intb2}
\end{equation}
where $b_{\rm max}$ depends on $v$ (see equation \ref{eqn:bmax}). Drawing a uniform random number $u \in [0,1]$ and solving $I(b) = u$ yields the analytical expression
\begin{equation}
b = b_{\rm max} \sqrt{u}.
\end{equation}

\clearpage
\bibliography{ms}{}
\bibliographystyle{apj}

\end{document}